\documentclass[12pt,preprint]{aastex}

\shorttitle{Formation and evolution of clumpy tidal tails around globular clusters}
\shortauthors{Capuzzo Dolcetta et al.}

\begin{document}


\title{Formation and evolution of clumpy tidal tails around globular clusters}


\author{R. Capuzzo Dolcetta}
\email{dolcetta@uniroma1.it}
\author{P. Di Matteo}
\email{p.dimatteo@uniroma1.it}
\and
\author{P. Miocchi}
\email{miocchi@uniroma1.it}
\affil{Dep. of Physics, Universit\'a di Roma La Sapienza, P.le Aldo Moro 2, 00185 Rome, Italy}


\begin{abstract}
We present some results of numerical simulations of a globular cluster orbiting
in the central region of a triaxial galaxy on a set of 'loop' orbits.
Tails start forming after about a quarter of the globular cluster orbital period
and develop, in most cases, along the cluster orbit, showing clumpy substructures
as observed, for example, in Palomar 5. If completely detectable, clumps can contain
about $7000 M_{\odot}$ each, i.e. about $10\%$ of the cluster mass at that epoch.
The morphology of tails and clumps and the kinematical properties of stars
in the tails are studied and compared with available observational data. Our
finding is that the stellar velocity dispersion tends to level off at large radii,
in agreement to that found for M15 and $\omega$ Centauri.
\end{abstract}




\section{Introduction}
Since Shapley's pioneering work \citep{shapley}, globular clusters (GCs) have played
a key-role in our understanding of the Universe and of the manner in which
our Galaxy formed: in the Milky Way they are the oldest stellar systems
found, with ages in the range $12$ to $15$ Gyr, so to represent tracers
of the early formation history of the Galaxy. \\
They are the best systems to study stellar dynamics,
having relaxation times smaller than their age, so that,
at least in the core, stars are expected to have lost memory of their
initial conditions \citep{BT}.
In the early 1980s a  number of approximated numerical studies of spherical self-gravitating systems \citep{cohn80} showed that the central density tends to increase dramatically over the time, so that ultimately a central power-law cusp is produced in the central region, even if these systems have an early evolutionary phase that resembles the King sequence of cluster models \citep{king66}.
Together with the slow collapse of the core, star evaporation occurs: the approach to equipartition implies that the
more massive stars sink toward the center of the cluster,
while lighter stars expand their orbit. Core collapse can be
halted by the presence of hard binaries which, acting as energy sources, heat
the central core by 3-bodies encounters \citep{hen61,ostr85}. \\
Internal processes are not the only responsible of dynamical evolution in
these systems: perturbations due to an external field (in particular, shocks due to the passage through
the galactic disk and to the interaction with the bulge) can accelerate significantly the
evolution of a globular cluster.
Indeed, it is commonly accepted that the present globular cluster population
represents the survivor of an initially more numerous one, depauperated by many disruptive processes \citep{mw97a,mw97b,fz01}. \\
There is observational evidence that the globular cluster system (GCS) radial profile is less peaked
than that of halo stars in our galaxy, M31 \citep{cv97}, M87 and M49 \citep{grill86, mcl95}, as well as in three galaxies of the Fornax cluster \citep{imma01} and of 11 elliptical galaxies \citep{cdt99}.
This fact leads to the hypothesis \citep{cdt97} that the two systems (halo and GCS) originally had the
same profile and that, afterwards, the GCS evolved mainly due
to two complementary effects: tidal interaction with the galactic field (which
causes less concentrated clusters to disintegrate more rapidly) and dynamical
friction (which induces massive globular clusters to decay in the central galactic
region in less than $10^{8}$ years, see \citet{cv03}).
External tidal fields have the effect of inducing the evolution of the shape of the mass
function of individual clusters, because of the preferential depletion of low-mass stars \citep{bm03} as a consequence of two-body relaxation . Strong evidence that the tidal field plays a fundamental role in the evolution of mass functions was
achieved by the discovery that their slopes correlate more strongly with the cluster location
in the Milky Way than with the cluster metallicity \citep{dpc93}.\\
In the last decade, many observational evidences of the interaction of GCs with the tidal field have been found.
Firstly,  \citet{grill95}, using colour-magnitude selected star counts in a dozen of galactic GCs,
showed that in the outer parts of these clusters the stellar surface density profiles exceeded the prediction of King models, extending also outside the tidal radius of the corresponding King model. Other results confirmed Grillmair's findings \citep{ls97, testa00, lmc00, sieg00, lee03}; all these works suggest that many GCs are likely surrounded by haloes or tails, made up of stars which were tidally stripped from the system. This was the state of the art until the spectacular findings of two tidal tails emanating from the outer part of the Palomar 5 globular cluster and covering an arc of 10 degrees on the sky, corresponding to a projected lenght of $4$ kpc at the distance of the cluster \citep{oden01,oden03}, obtained in the framework of the Sloan Digital Sky Survey Project (see also \emph{http://www.sdss.org}). \\
One of the relevant observational features of Palomar 5 is the presence of well defined clumps in the
star distribution along the tails \citep{oden01,oden03}. Also NGC 6254 and Palomar 12 seem to show clumpy structures in their tails \citep{lmc00}. This still deserves an exaustive interpretation. Actually, the  simulations of \citet{combes99} show
the presence of small clumps (containing about $0.5\%$ of the total number of stars of the cluster) in the tidal tails.
The authors attribute the formation of these clumps to strong gravitational shocks suffered by the cluster.
On another hand,  \citet{dehn04} were not able to reproduce the clumps observed in the tails of Pal 5,
even adopting a realistic galactic model, so that they argued that these structures could be due to the effect of
 Galactic sub-structures not accounted in their simulations (giant molecular clouds, spiral arms, dark-matter sub-halos
or massive compact halo objects).

With the aim of understanding better the
mechanism of interaction of GCs with the external field, in particular with
the bulge, and to investigate on the presence of clumps in tidal tails, we performed numerical simulations of globular clusters in orbit in  a triaxial galaxy, aiming also at
clarifying the morphological connection between the clusters tidal tails and their orbits.\\
In the next sections we show the results for globular clusters on 'loop' orbits in an inner region of a triaxial galaxy. In particular,
in Sect.\ref{Numericalmethods}, an overview of the numerical methods adopted to perform the
simulation are discussed; in Sect.\ref{Cluster} the galaxy and cluster  model adopted are presented; in Sect.\ref{results1} and Sect.\ref{results2} we deal with the main results of our work,
especially that concerning the formation of tidal tails around the cluster and their
orientation respect to the cluster orbit (Sect.\ref{tidal}), the radial density
profiles of the cluster, as they evolve with time, and the presence of clumpy regions in the tails (Sect.\ref{density}), the velocity
dispersion of stars in the cluster (Sect.\ref{vel}), the estimate of the
mass loss rate (Sect.\ref{loss}) and the evolution of the global mass functions (Sect.\ref{massf}); in the last section all the findings
are summarized and discussed.
\section{Numerical method}\label{Numericalmethods}
All the simulations were performed by means of an implementation of a
\emph{\/tree-code} carried out mainly by one of the authors
(P.M.). It is based on the algorithm described in \citet{bh} and adopts multipolar expansions
of the potential truncated at the quadrupole moment. It was parallelized
to run on high performance computers via MPI routines, employing an original parallelization approach \citep{bib2}.
The time-integration of the `particles' trajectories is performed by a
2$^{\rm{nd}}$ order leap-frog algorithm. This latter uses individual
and variable time-steps according to the block-time scheme \citep{aars,bib5},
in addition with corrections implemented in order to keep the same order of approximation
also during the time-step change.
The maximum time-step allowed is $\Delta t_{\rm{max}} =0.01 t_{\rm{c}}$ (where
$t_{\rm{c}}\sim (r^{3}_{\rm{core}}/GM)^{1/2}$ is the core-crossing time of the GC, being $M$ its
mass and $r_{\rm{core}}$ the core radius),
while the minimum is $\Delta t_{\rm{min}} = \Delta t_{\rm{max}} / 2^{8}$, thus fastest
particles may have a time-step as small as $\sim 4\times 10^{-5} t_{\rm{c}}$.
The best criterion we found for choosing the time-step of the $i$-th particle is via the formula:
$\min\{(d_i / a_i)^{1/2}, d_i / v_i\} / 20$, where $v_i$ is the velocity of the particle relative to
its first neighbour, $d_i$ the distance from its first neighbour and $a_i$ the modulus
of the acceleration.

To avoid instability in the time-integration, we smoothed the $1/r_{ij}$ interaction potential
by substiuting $1/r_{ij}$ with a continuous $\beta$-spline function
that gives an exactly Newtonian potential
for $r_{ij} > \epsilon$ and a force that vanishes for $r_{ij} \to 0$ \citep{bib5}.
In all the runs we set $\epsilon = 1.4\times 10^{-3}r_{core}$, so to have
$(\epsilon^3 / GM)^{1/2} \sim \Delta t_{\rm{min}}$.
Note that such value of $\epsilon$ is much less than the typical interparticle distance.
As regards the quality of the orbits time--integration, we checked that the
upper bound of the relative error in the energy conservation ($\Delta E/E$) is $10^{-8}$
per time--step, even in absence of the external field.

\section{Cluster and galaxy models}\label{Cluster}
\subsection{Galaxy model}\label{Galaxy}
The external galactic field due to the bulge is represented by the potential of the
Schwarzschild model \citep{scw}.
The Schwarzschild model is a non-rotating, self-consistent triaxial ellipsoid with axis ratios 2 : 1.25 : 1,
typical of many galaxies \citep{ber91}.\\
Defining adimensional units as
\begin{equation}
x^{\prime}=x/r_b,\ \ y^{\prime}=y/r_b,\ \ z^{\prime}=z/r_b, \\
\end{equation}
$r_b$ being the bulge core radius, the potential $\Phi(x^{\prime},y^{\prime},z^{\prime})$ is expressed as
the sum of a spherically simmetric term, $\Phi_{r^{\prime}}(r^{\prime})$, ($r^{\prime}=r/r_b$), which
corresponds to the potential  given by a density distribution following the modified Hubble's
law $\rho(r^{\prime})=\rho_0 \left[1+\left(r^{\prime}\right)^2\right]^{-3/2}$, ($\rho_0\equiv M_b/r_b^3 $),
plus two spherical armonics, $\Phi_1(z^{\prime},r^{\prime})$ and $\Phi_2(x^{\prime},y^{\prime},r^{\prime})$:\\ 
\begin{equation}
\Phi(x^{\prime},y^{\prime},z^{\prime})=A[\Phi_{r^{\prime}}(r^{\prime})+\Phi_1(z^{\prime},r^{\prime})+\Phi_2(x^{\prime},y^{\prime},r^{\prime})], \\ \label{pot_bul}
\end{equation}
where
\begin{eqnarray}
\Phi_{r^{\prime}}&=&-\frac{1}{r^{\prime}}\ln\left[r^{\prime}+\sqrt{1+\left(r^{\prime}
\right)^2}\right], \label {phir}\\
\Phi_1&=&c_1\frac{3\left(z^{\prime}\right)^2-\left(r^{\prime}\right)^2}{2(1+c_2\left(r^{\prime}\right)^2)^{3/2}}, \\
\Phi_2&=&-3c_3\frac{\left(x^{\prime}\right)^2-\left(y^{\prime}\right)^2}{(1+c_4\left(r^{\prime}\right)^2)^{3/2}},\\
A&=&4\pi\frac {G M_b}{r_b},
\end{eqnarray}
and $M_b$ is the bulge mass.\\
The coefficients $c_i$ have the values: $c_1=0.06408$, $c_2=0.65456$, $c_3=0.01533$, $c_4=0.48067$ \citep{zeeuw,pesce}.
They have been determined so to have density axial ratios roughly constant with $r$. \\
Following \citet{pesce}, we will consider $M_{b}=3\times 10^{9}$ M$_{\odot}$ and $r_{b}=200$ pc, but the
results obtained in adimensional variables (see Appendix \ref{scal}) are scalable
in terms of $r_b$ and $M_b$, for given initial conditions for positions, velocities and
$m_i/M_b$.

\subsection{Cluster model}\label{Cluster1}
As initial cluster model, we chose a multimass King distribution \citep{king66,costa76}, with $10$
mass classes, ranging between $0.12$ and $1.2 $M$_{\odot}$ and equally spaced in a logarithmic scale.
To find the distribution function for each mass class, we integrated the Poisson's
equation as described in the Appendix \ref{kingmod}.\\
The initial cluster
mass function is chosen in the Salpeter form \citep{salp}, i.e. $dN/dm \propto m^{-2.35}$
(see Appendix \ref{remn} for a discussion about the remnants of progenitor stars more massive
than $1.2 M_{\odot}$). We included mass
segregation in the initial conditions of our cluster model because we wanted to simulate a
dynamically-relaxed cluster
(supposed to be sufficiently massive to have frictionally decayed in the central galactic region).
The ``initial'' mass of the system is $M_{tot}=3\times10^{5}$ M$_{\odot}=10^{-4}M_b$, the initial
concentration parameter is $c=log(r_{t}/r_{\rm{core}})=1.1$ and the central velocity dispersion is
$\sigma=9.4$ km s$^{-1}=0.036\times r_b/t_{cross}$, being $t_{cross}=(r_b^3/GM_b)^{1/2}$ the bulge
crossing time.\\
Then, the system was represented by a number ($N$) of `particles' lower than the number of stars
in the real cluster, with masses properly rescaled such to give a total mass equal to $M_{tot}$.
The cluster moves on the y-z coordinate plane, following loop orbits of different ellipticity
\begin{equation}
e=\frac{R_a-R_p}{R_a+R_p},\label{ell}
\end{equation}
being $R_a$ and $R_p$, respectively, the apocenter and pericenter distances (see Table \ref{tbl-1}
for orbital parameters).



\section{Results on tidal tails}\label{results1}

In the following, we present the main findings of our work concerning with tidal tails structure and
evolution.
When we refer to the center-of-density of the cluster, we mean a mass density-weighted center as defined
by \citet{ch85}.

\subsection{Tidal tails formation and morphology}\label{tidal}

\begin{figure}
\plotone{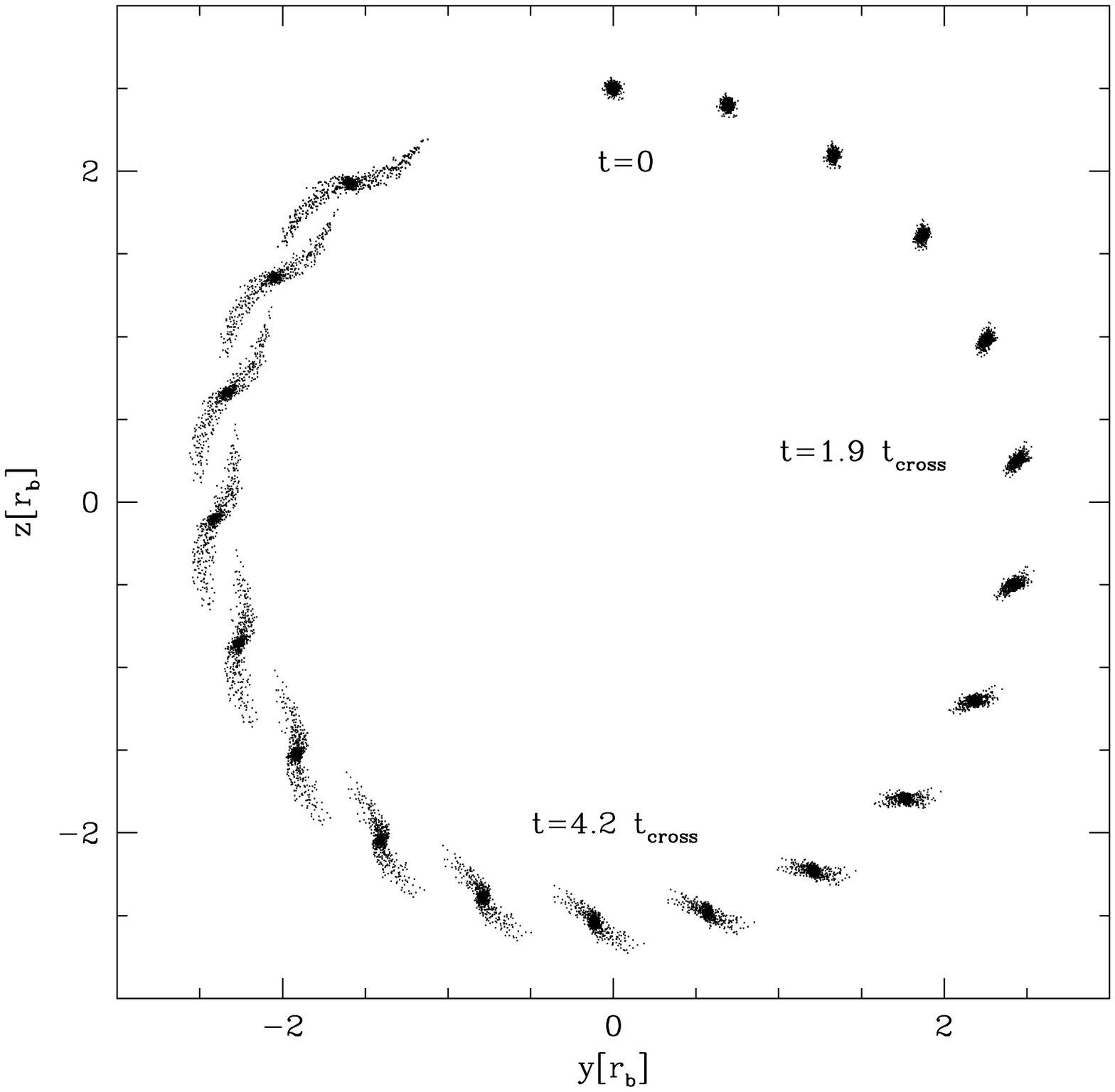}
\caption{First orbital period of a $3\times 10^{5}M_{\odot}$ globular cluster in the
potential described in Sec. 2.2. The cluster moves on a quasi-circular orbit ($e=0.03$)
around the galaxy center in a clockwise direction (see text). Distances are in units
of the galactic bulge radius $r_b$. Some snapshots are labelled with time, expressed
in units of the galactic bulge crossing time $t_{cross}$.\label{orb1}}
\end{figure}

\begin{figure}
\plotone{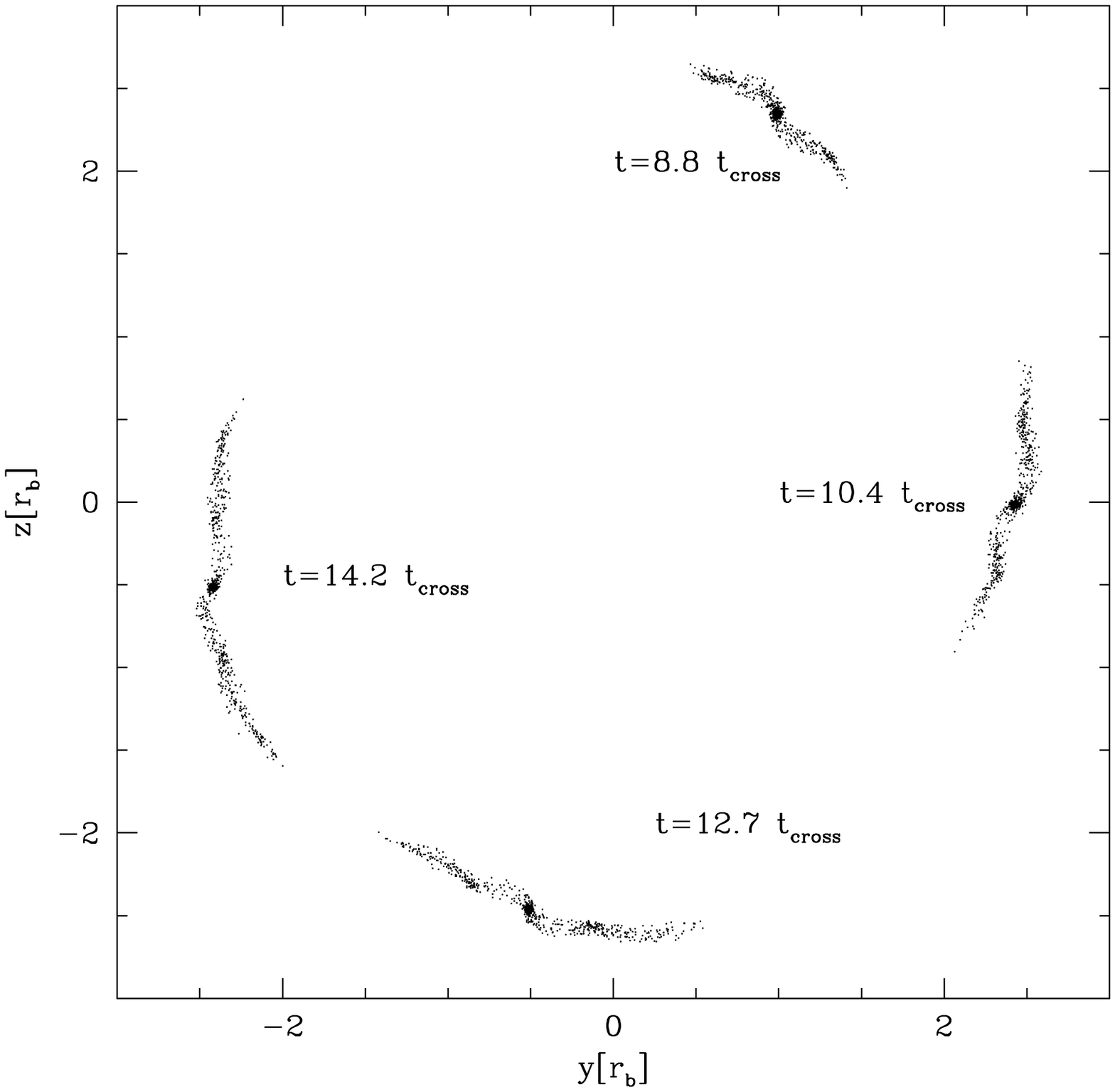}
\caption{Second orbital period of the globular cluster described in the Fig.\ref{orb1}.\label{orb2}}
\end{figure}

\begin{figure}
\plotone{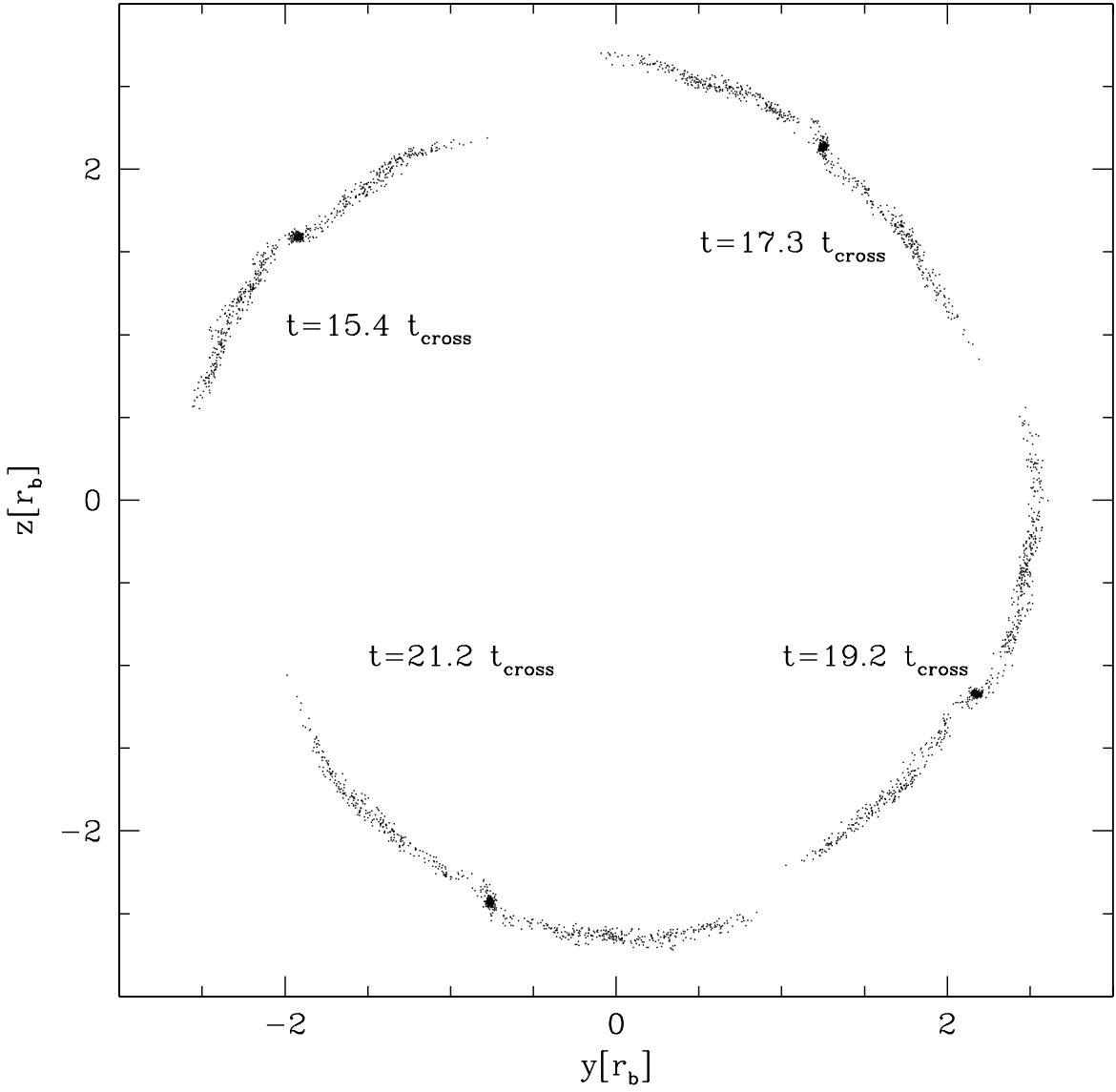}
\caption{Third orbital period of the globular cluster described in the Fig.\ref{orb1}. \label{orb3}}
\end{figure}

\begin{figure}
\plotone{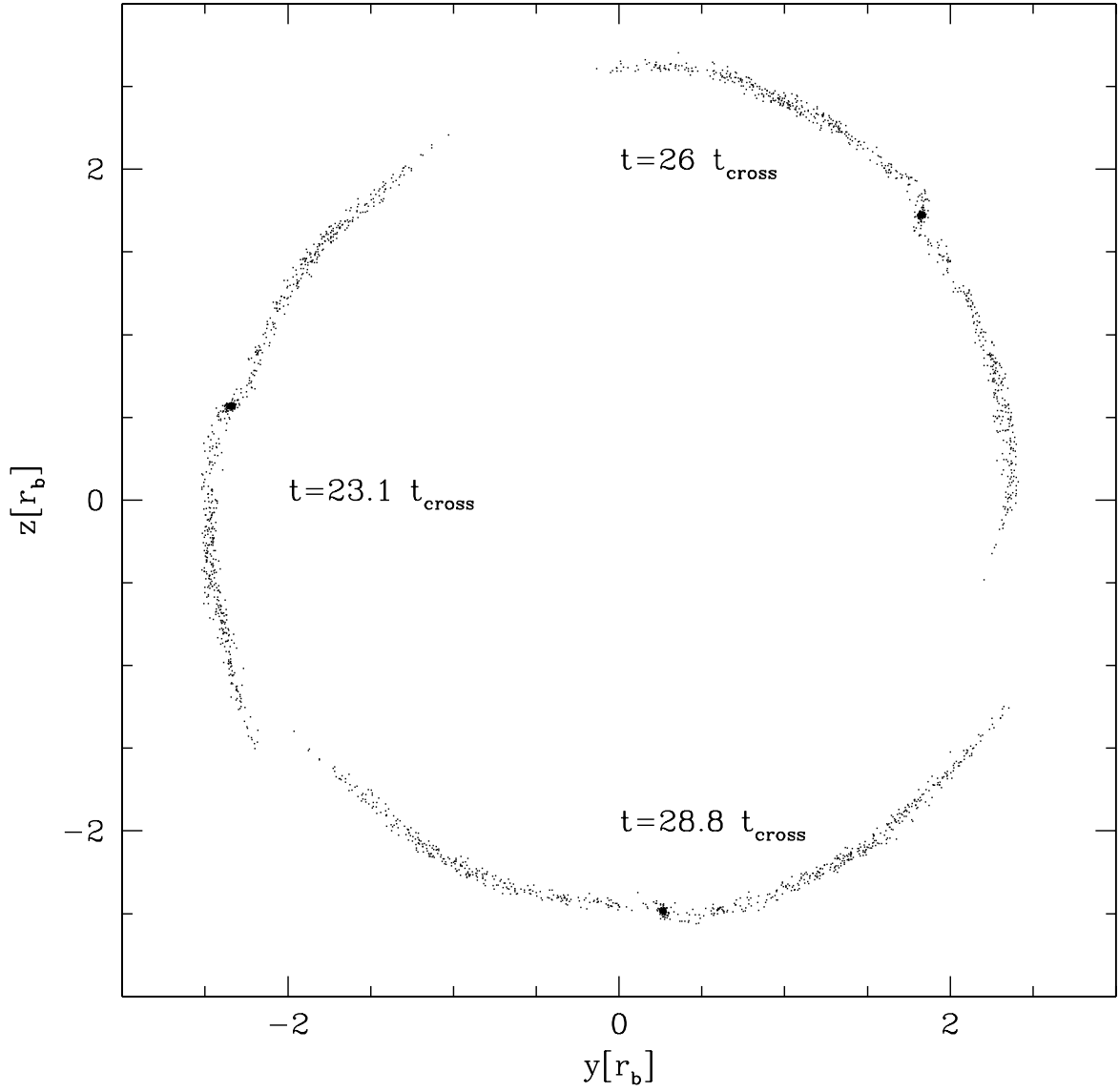}
\caption{Last orbital period of the globular cluster described in the Fig.\ref{orb1}. \label{orb4}}
\end{figure}

\begin{figure}
\plotone{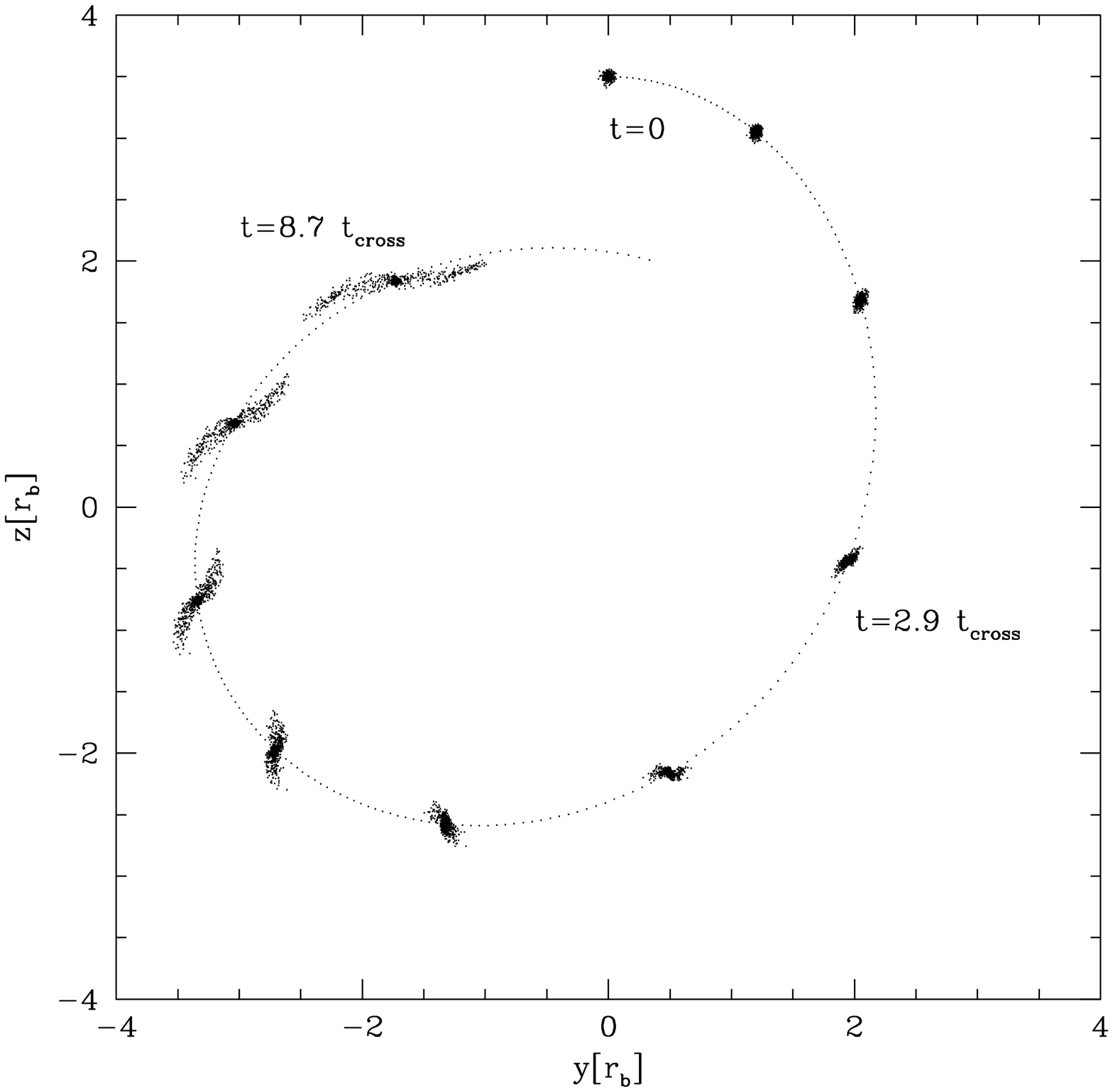}
\caption{First orbital period of the $3\times 10^{5}M_{\odot}$ globular cluster in a loop
orbit with ellipticity $e=0.27$ around a triaxial galaxy. The cluster moves in a
clockwise direction. Some snapshots are labelled with time. The dotted line represents
the cluster orbit.  \label{orb5}}
\end{figure}

\begin{figure}
\plotone{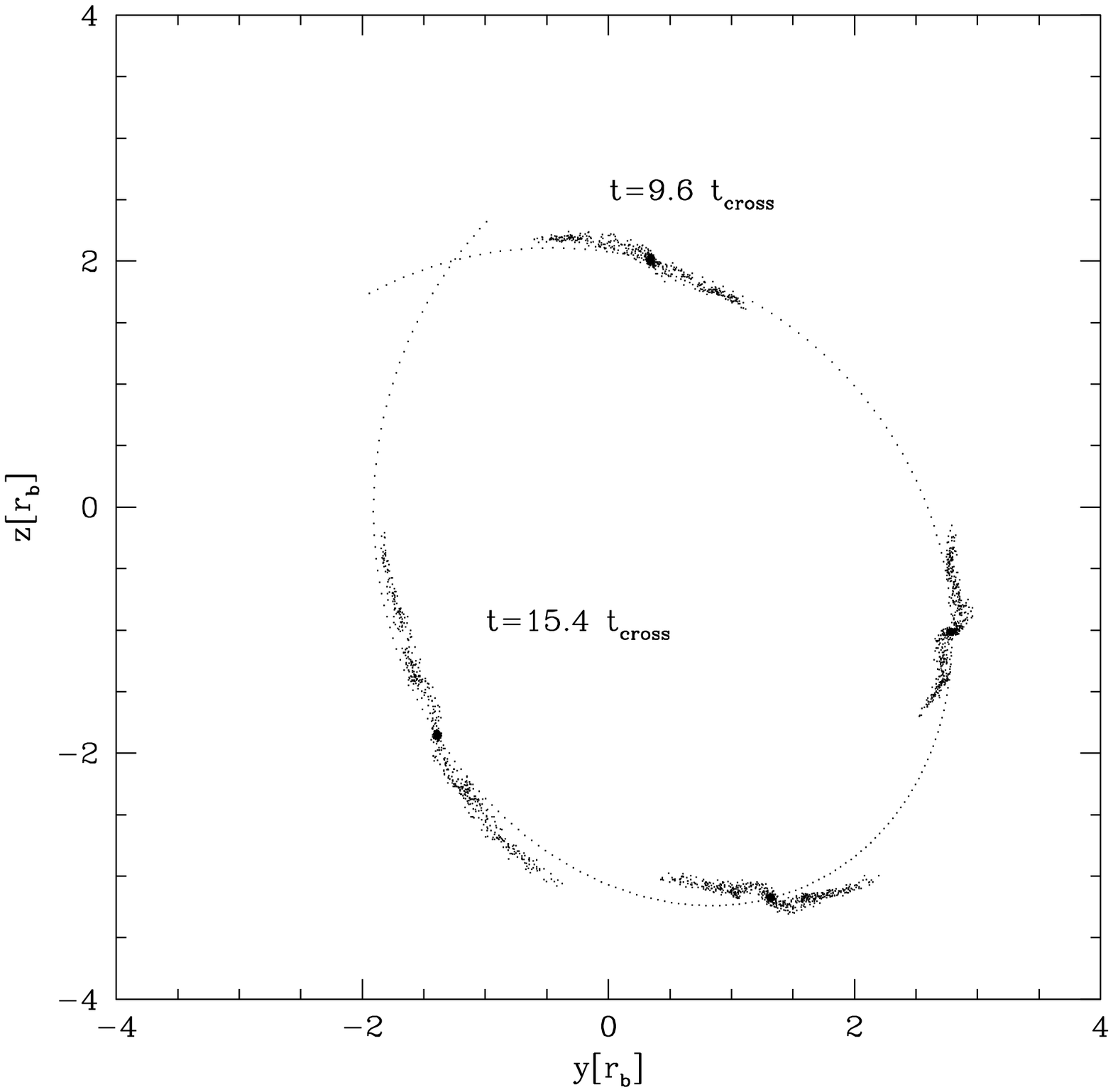}
\caption{Second orbital period of the globular cluster described in the Fig.\ref{orb5}. \label{orb6}}
\end{figure}

\begin{figure}
\plotone{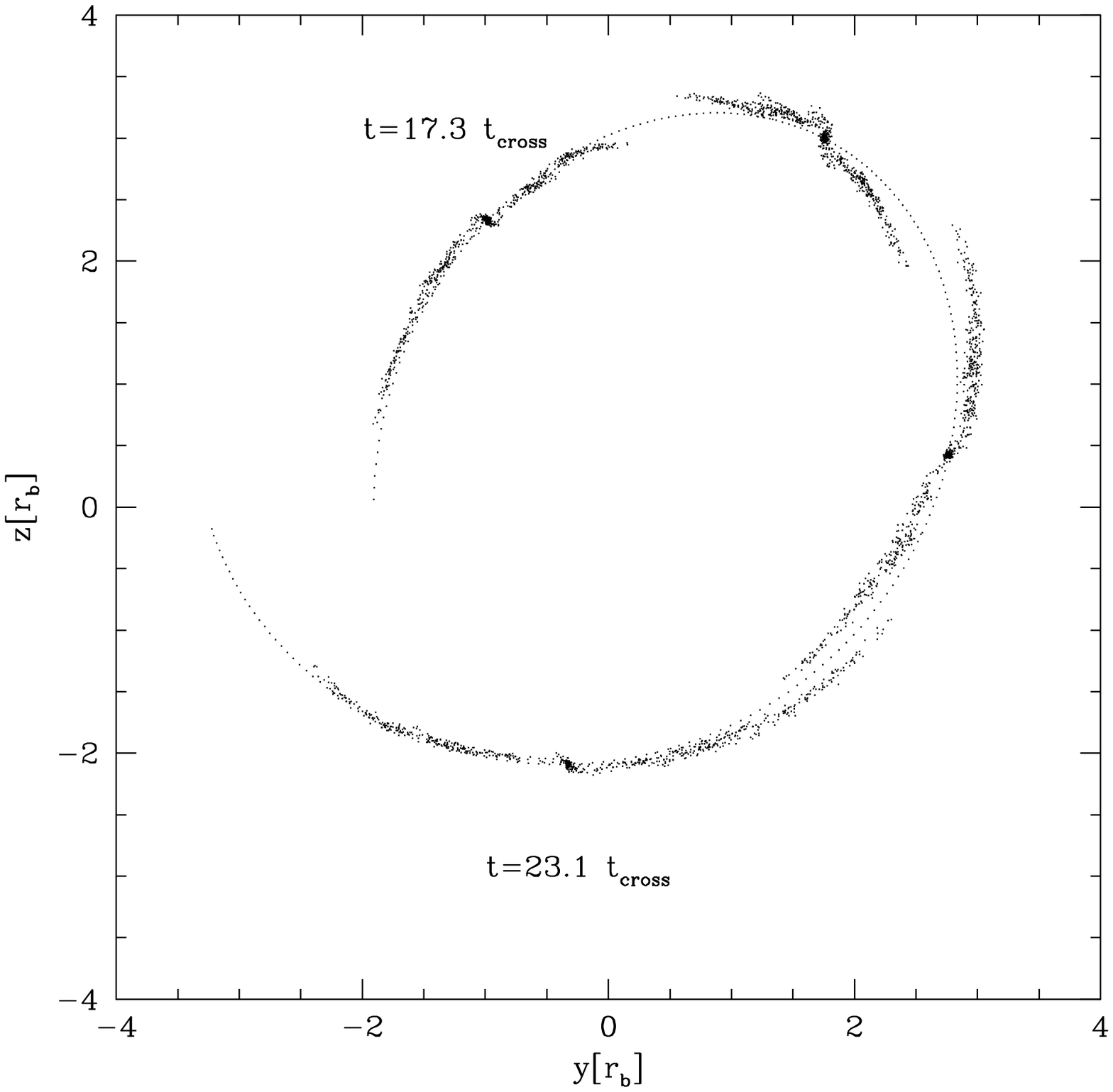}
\caption{Third orbital period of the globular cluster in the Fig.\ref{orb5}.\label{orb7}}
\end{figure}

\begin{figure}
\plotone{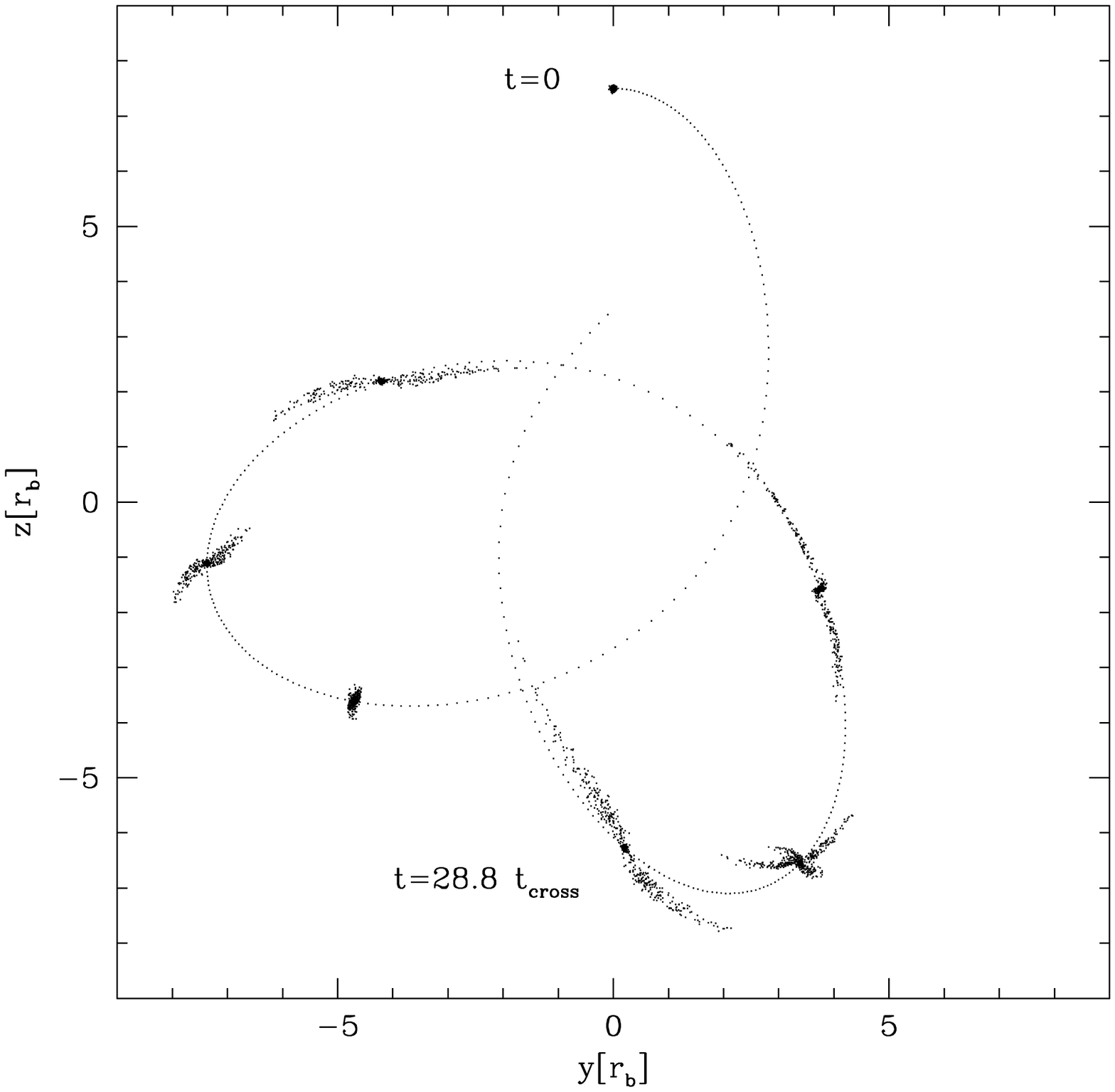}
\caption{Snapshots of the $3\times 10^{5}M_{\odot}$ globular cluster in a loop orbit with
ellipticity $e=0.57$ around a triaxial galaxy. The cluster moves in a clockwise direction.
Some snapshots are labelled with time. The dotted line represents the cluster orbit.\label{orb8}}
\end{figure}

\begin{figure}
\plotone{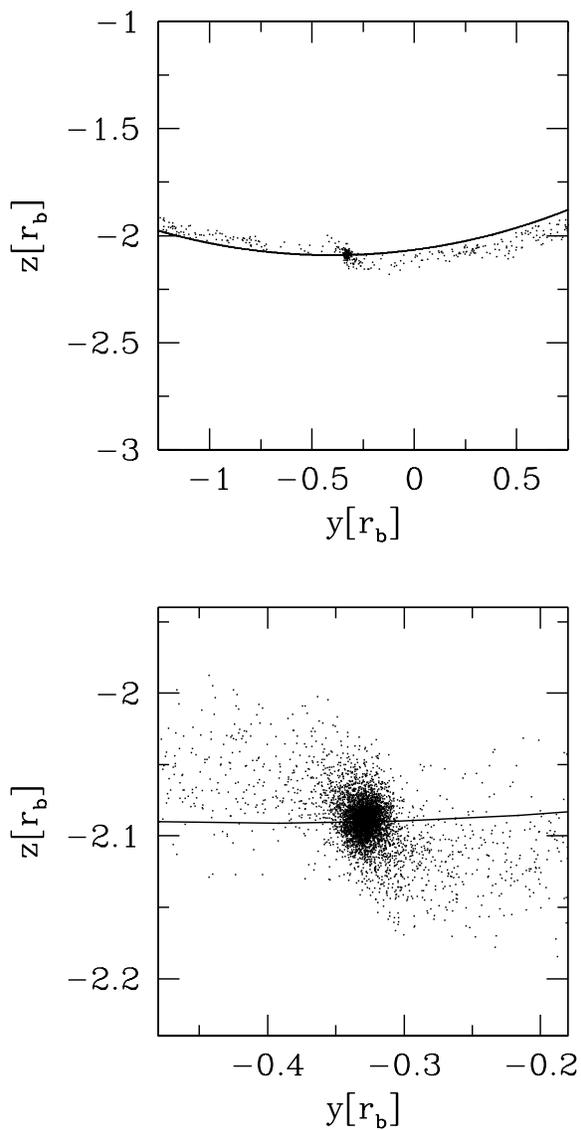}
\caption{Snapshot of the $3\times 10^{5}M_{\odot}$ globular cluster in the loop orbit with
$e=0.27$ at t=23.1$t_{\rm{cross}}$. The upper panel shows the system and the orbit described
by the cluster density center (solid line). It is evident from the zoom in the bottom panel
that the tails around the cluster core can lead to not reliable information about the orbital
path of the cluster (solid line). \label{zoom}}
\end{figure}

In all the simulations performed, the cluster starts moving around the galaxy center in a clockwise direction
(seen from the positive x axis). The different loop orbits have been followed for about 30 $t_{\rm{cross}}$.\\
In Fig.\ref{orb1}, \ref{orb2}, \ref{orb3}, \ref{orb4}, \ref{orb5}, \ref{orb6}, \ref{orb7} and \ref{orb8}, the formation and subsequent
development of tails around the globular cluster is shown.\\
After about 8 $t_{\rm{cross}}$, tidal tails are clearly formed. They
continuosly accrete by stars leaving the cluster, so that after 30 $t_{\rm{cross}}$, in the case of
quasi-circular orbit ($e=0.03$), they are elongated for more than 3 $r_b$ each and contain about $75\%$
of the initial cluster mass.
As it is clearly visible from these figures, the degree of elongation of the tails along the cluster orbital
path strongly depends on the ellipticity $e$ (Eq.\ref{ell}) of the cluster orbit.
Indeed, while in the case of the quasi-circular orbit tails are a clear tracer of the cluster path, in
the most eccentric orbit ($e=0.57$), tails are strictly elongated along the orbital path only when the
cluster is near the perigalacticon, while at the apogalacticon they tend to deviate from the
cluster path.
Nevertheless, in \citet{mioc} a remarkable tails---orbit alignment is found for clusters moving on
quasi--radial orbits in the same bulge potential.
However, it is important to stress that, in order to perform accurate predictions of the cluster
orbit from observational detections of tidal tails, it is necessary to look at the spatial distribution
of stars well outside the cluster (typically $2-3$ times the cluster limiting radius).
Indeed, in the vicinity of the globular cluster, stars in the tails are not aligned with the cluster orbit,
neither in the case of small ellipticity (see Fig.\ref{zoom}), but they distribute along the peculiar
$S-$shape profile not aligned along the orbit.\\
The orbit ellipticity also influences the similarity between the two cluster tails.
For the quasi-circular orbit, these structures are simmetric for the whole duration of the simulation,
being elongated, at a given time, for the same length. For more eccentric orbits, the leading tail tends
to be more elongated than the trailing one when going from the apocenter to the orbital pericenter and,
viceversa, it is less elongated than the trailing tail when the cluster moves towards the apocenter.
In any case, the tail that precedes the cluster extends always slightly below the orbit
while the trailing one lies slighlty above this latter,
in agreement with what observed for Palomar 5. \\
The shape and orientation of the tails can be easily understood in the case of a cluster moving
on a circular orbit in an axysimmetric external field, using a rotating frame of reference with the
origin in the baricentre of the cluster, with the $X$-axis pointing towards the galactic center,
the $Y$-axis parallel to the direction of motion of the cluster and the $Z$-axis orthogonal to
the orbital plane. In this reference frame, the galactic tidal field tends to accelerate stars
along the $\pm X$ directions \citep{hh03}, making stars to escape from the system through the
Lagrangian points $L_1$ and $L_2$ (which are the two equilibrium points located along the $X$-axis).
But the Coriolis acceleration tends to align escaping stars along the direction of motion of the
cluster around the galaxy, this yelding the peculiar $S$-shape just outside the cluster,
in the inner part of the tails.\\


\subsection{Density profiles}\label{density}
\begin{figure}
\plotone{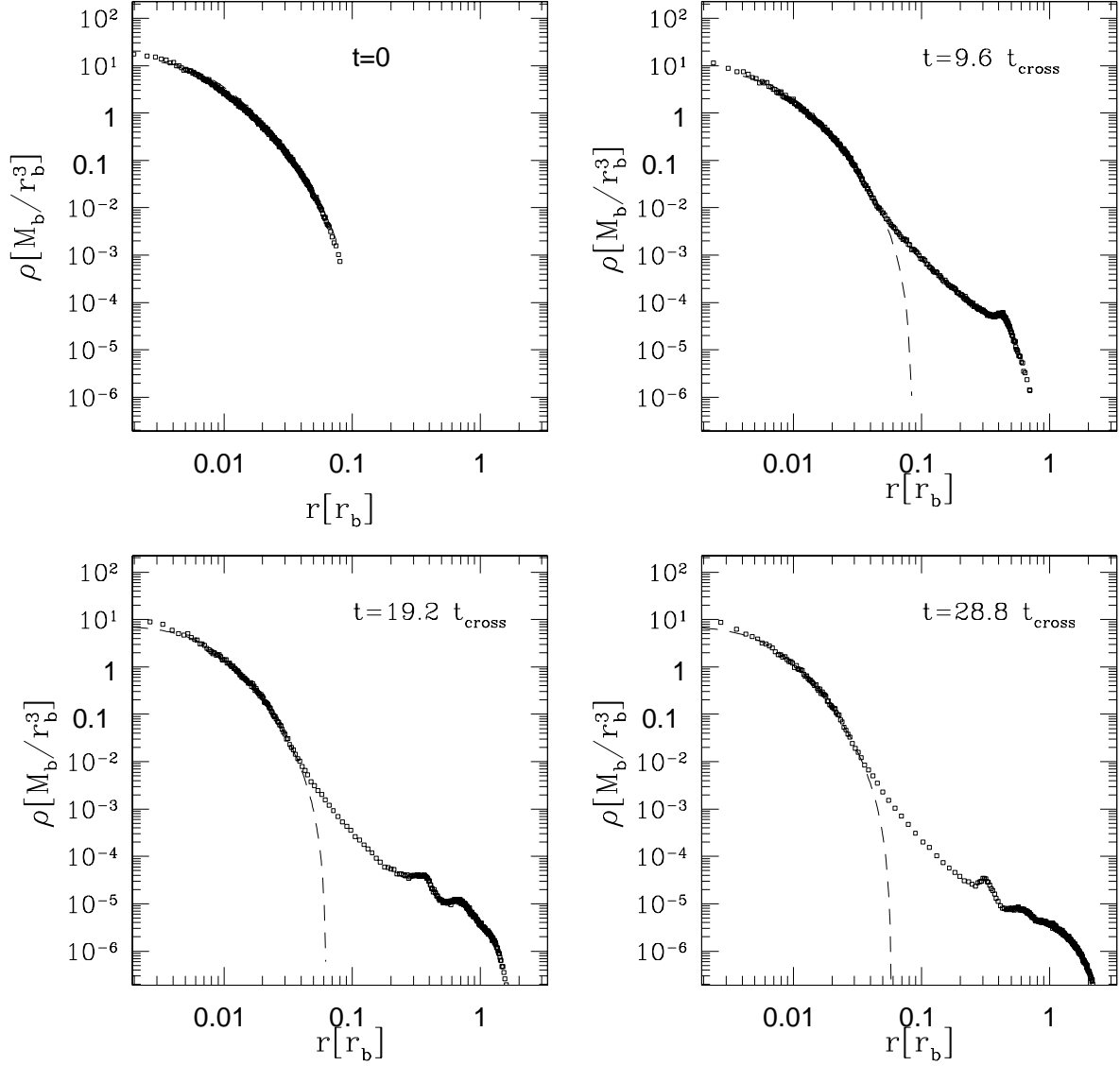}
\caption{Volume mass density of the cluster in the case of the quasi-circular orbit ($e=0.03$),
at four different epochs, as labelled. The dashed line in each panel represents the best King
model fit at that epoch. The presence of clumps in the tails are clearly visible.\label{voldens}}
\end{figure}

\begin{figure}
\plotone{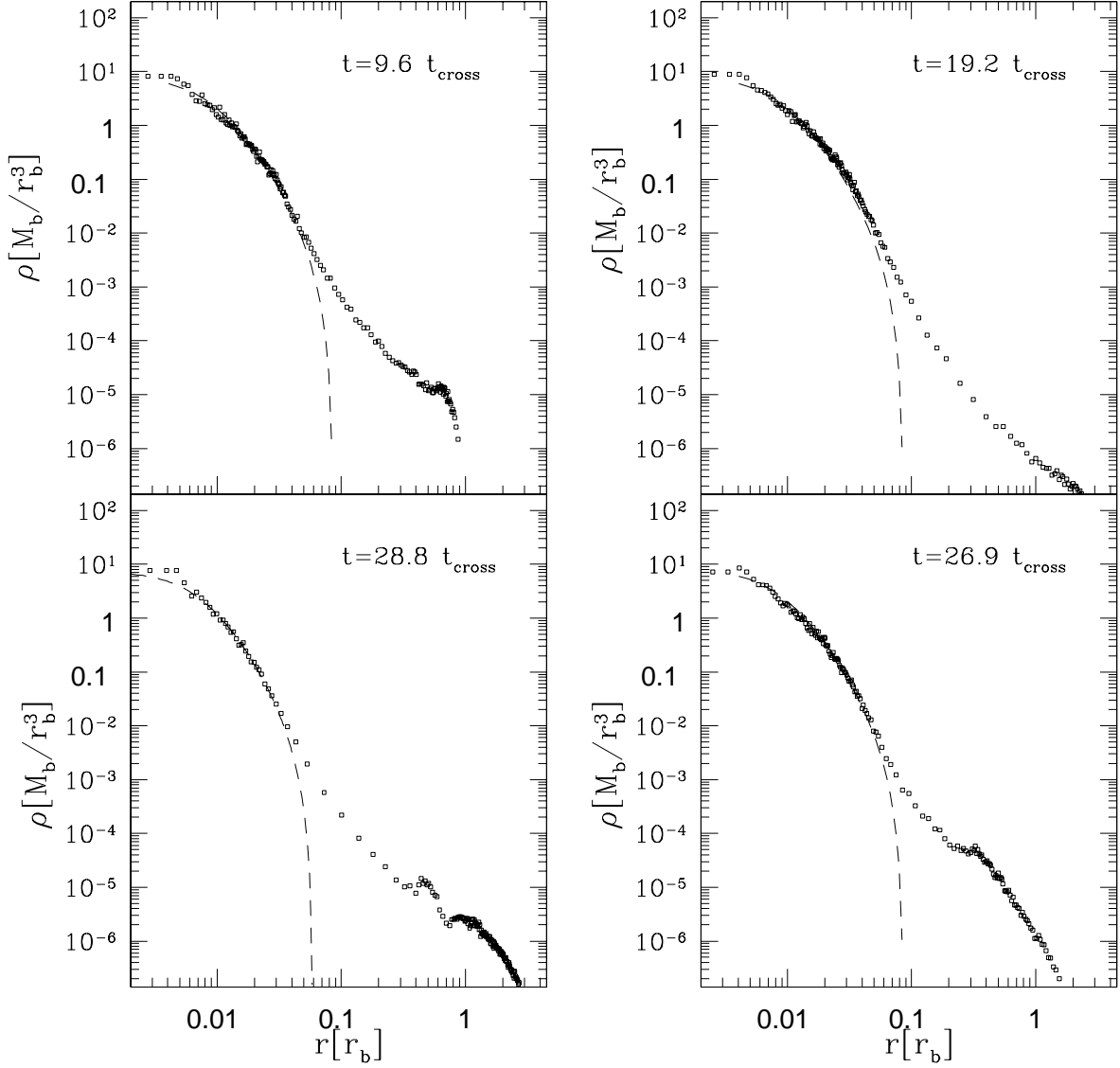}
\caption{Volume mass density of the cluster moving on the loop orbits with $e=0.27$ (left column)
and $e=0.57$ (right column), at different epochs. The dashed line in each panel
represents the best King model fit at that epoch. Note that in the case of most eccentric
orbit, clumps are not yet formed at t=19.2$t_{\rm{cross}}$. \label{voldenstutte}}
\end{figure}

\begin{figure}
\plotone{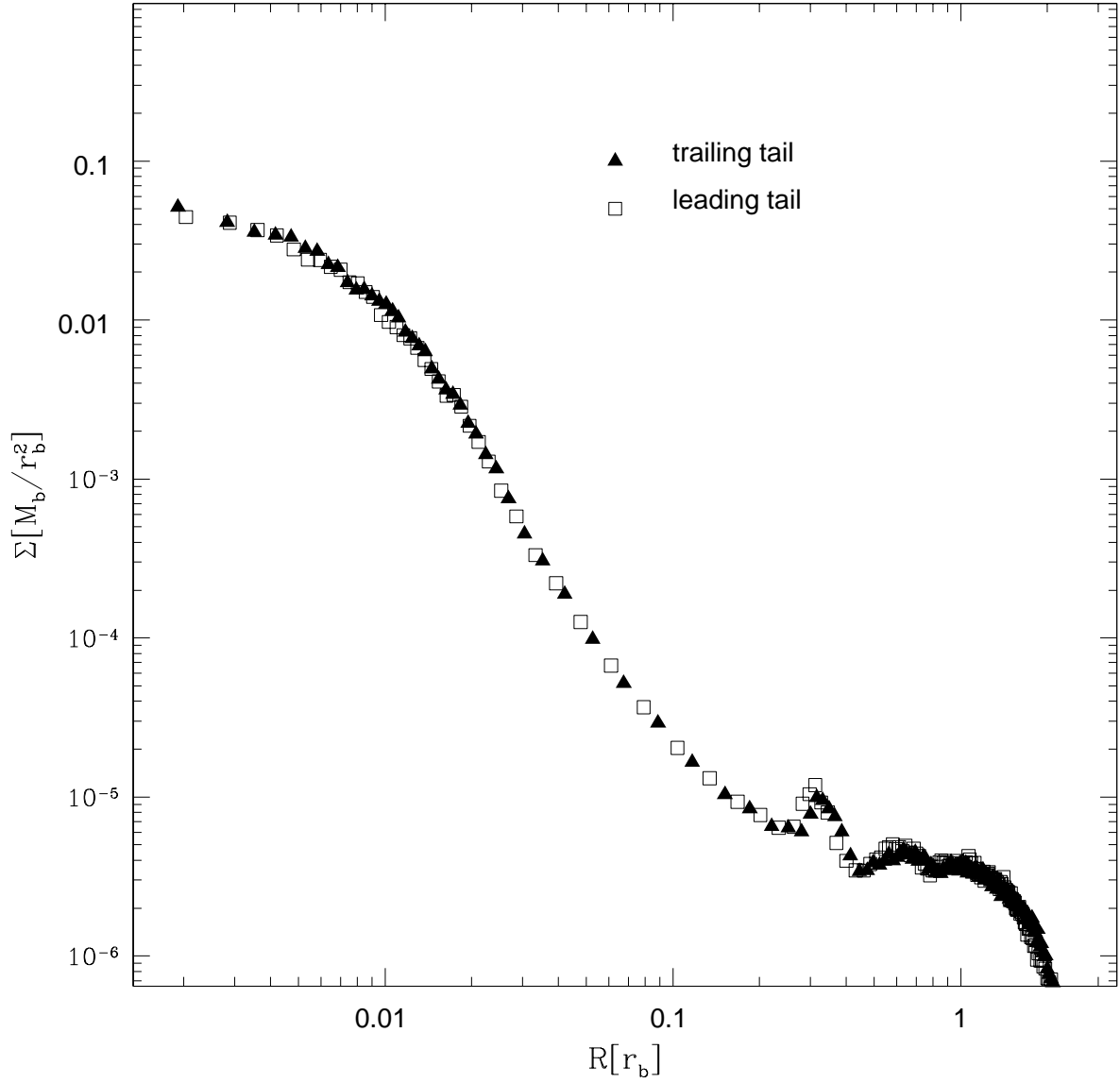}
\caption{Surface density profile of the cluster at t=28.8$t_{\rm{cross}}$ in the case of
the quasi-circular orbit. Two different regions are plotted: that containing the trailing
tail (filled triangles) and that containing the leading tail (open squares).
The line-of-sight is parallel to the x axis and so perpendicular to the orbital
plane.\label{supdens2}}
\end{figure}

\begin{figure}
\plotone{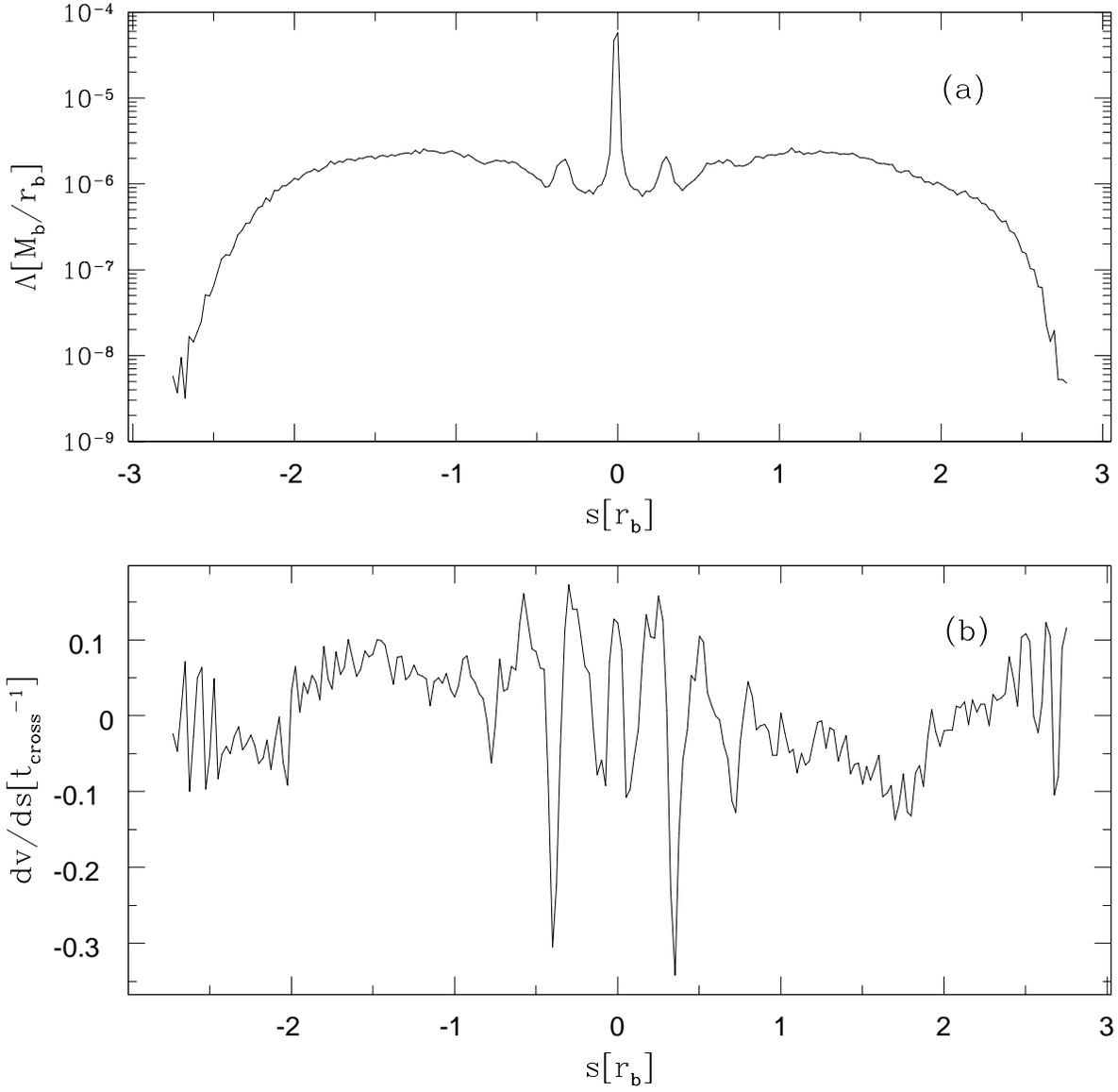}
\caption{Panel (a): Linear mass density as a function of the curvilinear abscissa $s$,
set equal to zero at the cluster center, negative for the traling tail and positive
for the leading tail.
Panel (b): Derivative of the stellar tangential velocity respect to $s$.  \label{teta}}
\end{figure}

In order to describe the tidal debris and to compare our findings with observations
\citep{ls97,testa00,lmc00,sieg00,lee03,oden03}, we studied the radial profile of the volume
and surface densities
(azimutally averaged) as a function of the distance from the cluster center.
Obviously, this description does not take into account the fact that stars lost from the cluster are
not placed in a spherically symmetric structure, but it has the advantage to provide a global
study of both the cluster and the tails that can be easily compared with observational data.
In Fig.\ref{voldens} and \ref{voldenstutte}, the volume density of the system is shown at different epochs,
for the various orbits.  Once the tails have completely developed, outside the \emph{S}-shape distribution,
density clumps appears. They are symmetrically located in the two tails, as shown in Fig.\ref{supdens2} for
the cluster on quasi-circular orbit: in this case, the most prominent clumps  are located at a distance
from the cluster center between $0.25r_b$  and $0.4r_b$. The density profiles  are very similar to that
of Palomar 5, where clumps are visible in the outer part of the cluster \citep{oden03}.\\
Of course, the possibility to detect observationally these clumps is strongly related to the cluster
position along its orbit.
Indeed, we computed the contrast density ratio $\rho_{cl}/\rho_{\ast}$, where $\rho_{cl}$ is the local
volume density in the clumps and $\rho_{\ast}$ is the background (i.e. the bulge) density around them.
This ratio is maximum when the cluster is near apogalacticon and decreases when moving towards perigalacticon, as shown in Table \ref{tbl-2}. This is due to two complementary effects: when the cluster is near apogalacticon, $\rho_{\ast}$ is minimum (according to the galaxy model described in Sect.\ref{Galaxy}) and, at the same time, the elongation of the tails tends to compress respect to that at perigalacticon (see, for example, Fig.\ref{orb7} and Fig.\ref{orb8}) and so $\rho_{cl}$ increases. If completely detectable, clumps can contain about $7000 M_{\odot}$ each (i.e. about $10\%$ of the cluster mass at that epoch),
as in the case of the cluster moving on the quasi-circular orbit after $30 t_{cross}$.\\
In order to study the mass distribution along the tails, we have also evaluated the ``linear'' density for the whole system in the quasi-circular orbit around the galaxy.
This study is particularly well fitted to investigate the mass distribution because tails form
a long and thin structure.
The upper panel of Fig.\ref{teta} shows the linear mass density as a function of the curvilinear abscissa
$s$ along the system: the absolute maximum in the plot corresponds to the cluster location, while the two
simmetric relative maxima correspond to the two main clumps. These clumps result to be unbound structures
(see also \citet{dm04});
we followed the motion of stars that at a certain time stay in the two clumps: they crowd in the clumps
for some time and then move away in the outer parts of the tails.  Once moved away from clumps,
these stars tend to disperse along the cluster tails. Also the simmetric location of these two
clumps respect to the cluster center makes improbable that these structures can be due to
local disomogenities in the gravitational field along the tails. More probably, these clumps
are related to cinematical properties of stars in their surroundings. The bottom panel
in Fig.\ref{teta} shows the derivative of the stellar tangential velocity component with respect
to the curvilinear
abscissa $s$ defined above. As is evident, the two clumps correspond to a region where this derivative
has a negative minimum, which is also the global minimum over the whole extension of the tails.
This implies that the local velocity of the stars decreases as they are approaching the clumps, thus
leading to the local overdensity which originates such structures.
However, the mechanism at the basis of the formation of these structures still requires further and
more detailed investigationsd that we postpone to next papers. See, however, the discussion in
\citet{mioc} for the case of clusters in quasi--radial orbits.

\subsection{Velocity dispersion}\label{vel}
\begin{figure}
\plotone{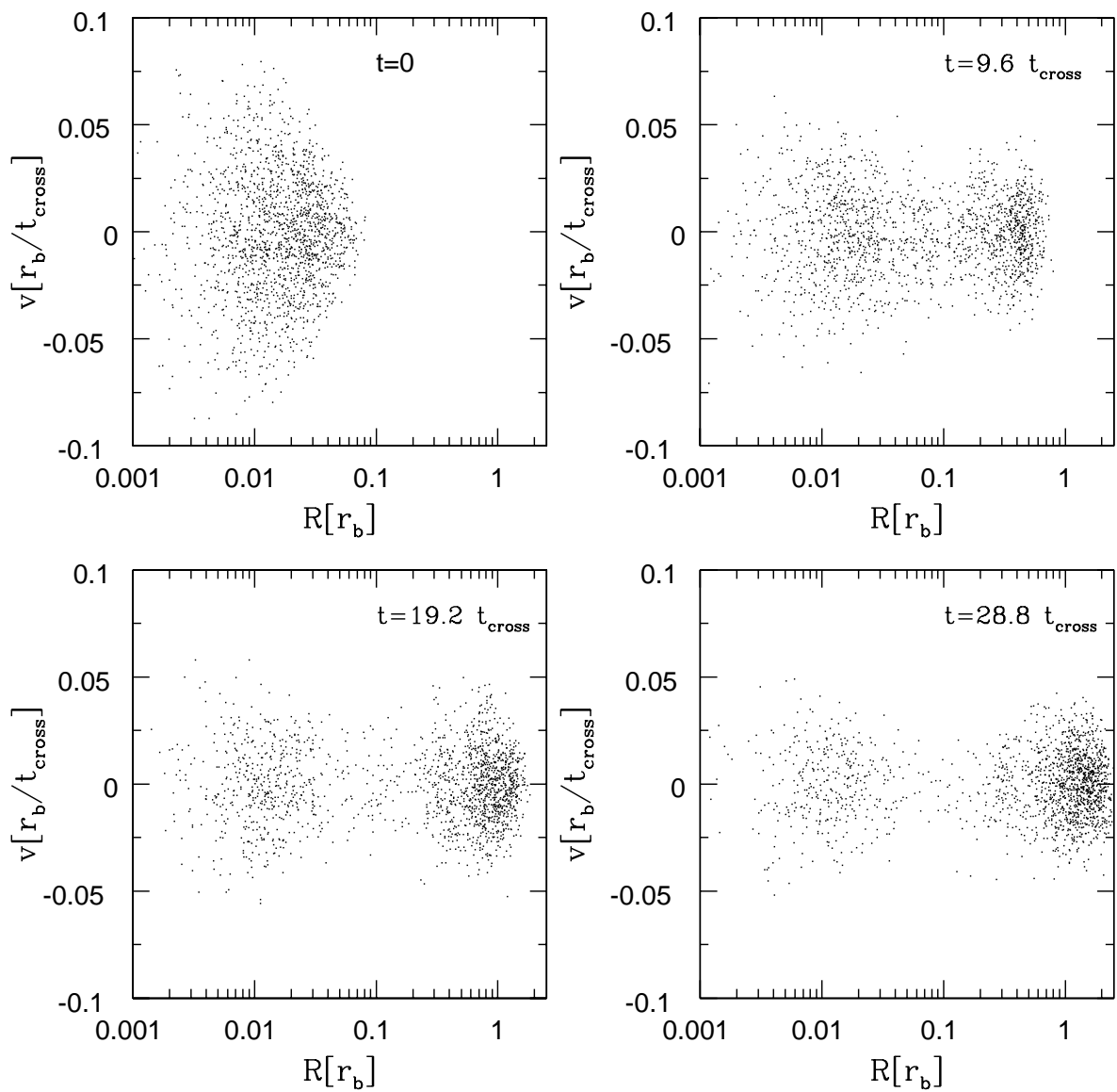}
\caption{Stars velocities along the x axis vs. the distance from the cluster center,
for the case of the quasi-circular orbit ($e=0.03$), plotted at four different epochs.
Once stars begin to escape from the cluster, the velocity profile shows a minimum and
then increases again. \label{vlos}}
\end{figure}

\begin{figure}
\plotone{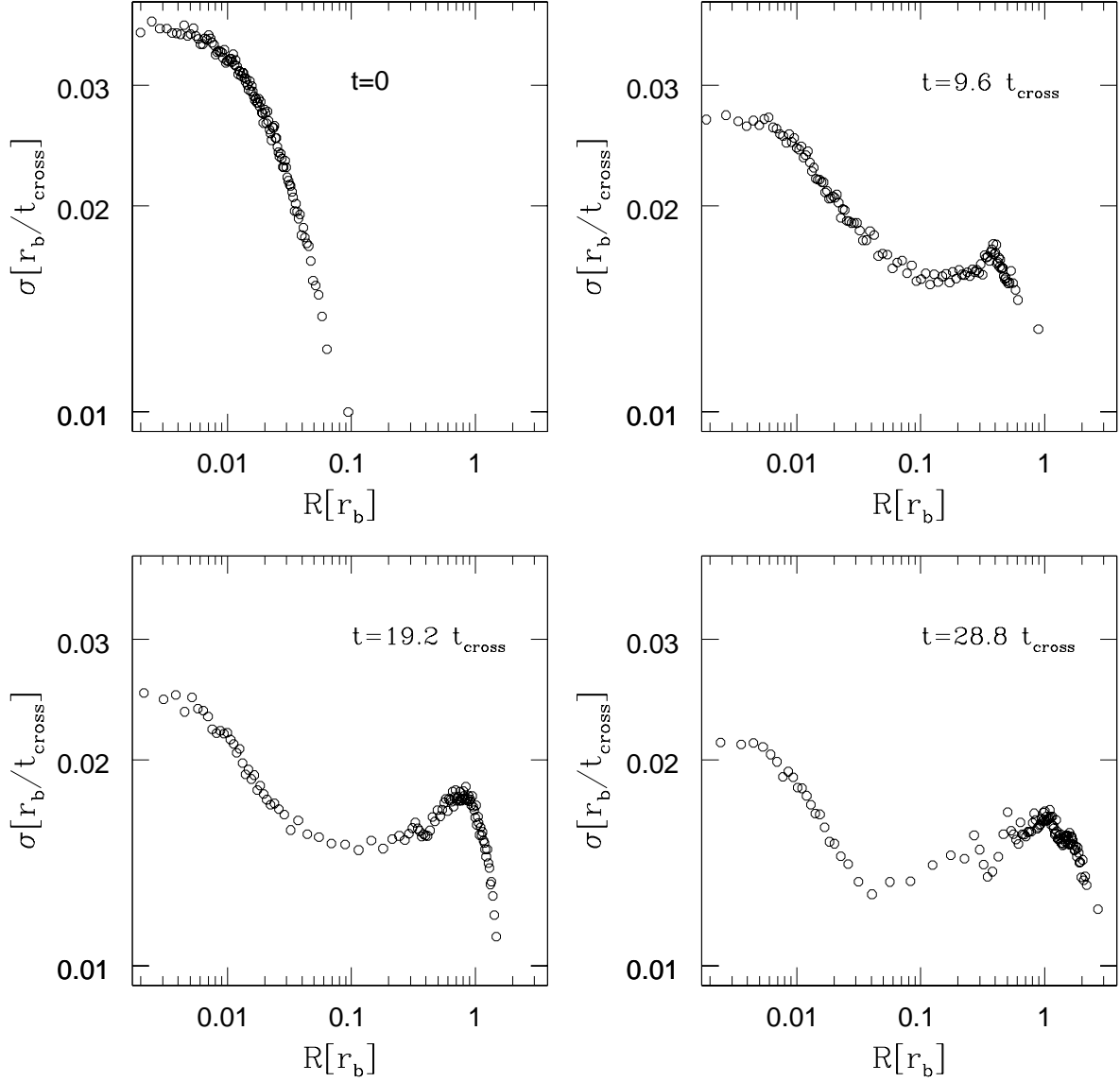}
\caption{Velocity dispersion profiles along the x axis for the cluster in the
quasi-circular, at four different epochs.\label{disp}}
\end{figure}

For the two galactic globular clusters M15 and $\omega$ Centauri there are observational evidences
that the stellar velocity dispersion remains constant at large radii \citep{scarpa,druk}.\\
Three hypotheses have been raised to justify these findings: 1) tidal heating, as suggested by Drukier for M15 \citep{druk}; 2) the presence of a dark matter halo surrounding the clusters \citep{car00}; 3) a breakdown of Newton's law of gravity in the weak acceleration regime \citep{scarpa}.\\
We studied the velocity dispersion profile of our simulated cluster as it would be detected if the
system was seen along a line-of-sight perpendicular to the cluster
orbital plane.
In Fig.\ref{vlos}, line-of-sight velocities of members of the cluster are plotted versus distance
from the center, at four different epochs. At $t=0$, the velocities decrease moving from the center of
the cluster outwards, as it is expected from a King model with a tidal cutoff.
As the system moves through the galaxy and loses stars, the velocity profile varies significantly: it tends
to decrease until a limiting value and then increases again. This behaviour is very similar to that
found in M15 (cfr Fig.8 in Drukier et al. 1998). \\
In Fig.\ref{disp} the line-of-sight velocity dispersion profile is shown. It is evident from the Figure
that in the outer part of the cluster, the dispersion tends to level off. This region corresponds to that characterized by a power-law volume density profile (see Fig.\ref{voldens} and \ref{voldenstutte}).
Stars in this region are escaping from the system and their motion is mostly oriented along the radial direction towards
the galaxy center. Once escaped, they move around the galaxy weakly interacting with each other, with
similar orbital parameters, so that the velocity dispersion found is
coherent with that of a set of particles moving in the triaxial potential adopted.\\
The second relevant finding is the decreasing of the velocity dispersion in the inner part of the cluster,
which could be explained by the quick revirialization of the inner part as stellar mass is being lost.
This in accordance with the very low velocity dispersion of Pal5 \citep{oden02}, a cluster which has
suffered a great mass loss, as it is now well estabilished.

\section{Results on mass loss}\label{results2}
\subsection{Mass loss}\label{loss}
To estimate the mass loss from the cluster, we decided to use an `observational' definition.
At any given time we compare the cluster local density $\rho_{gc}$ with the background stellar
density $\rho_{\ast}$, assuming that a star is actually belonging to the cluster if it is located
in a region dense enough to make it distinguishable from the background, i.e. if
\begin{equation}
\frac{\Delta\rho}{\rho_{\ast}}\geq 1,
\end{equation}
being
\begin{equation}
\Delta\rho=(\rho_{\ast}+\rho_{gc})-\rho_{\ast}.
\end{equation}

\begin{figure}
\plotone{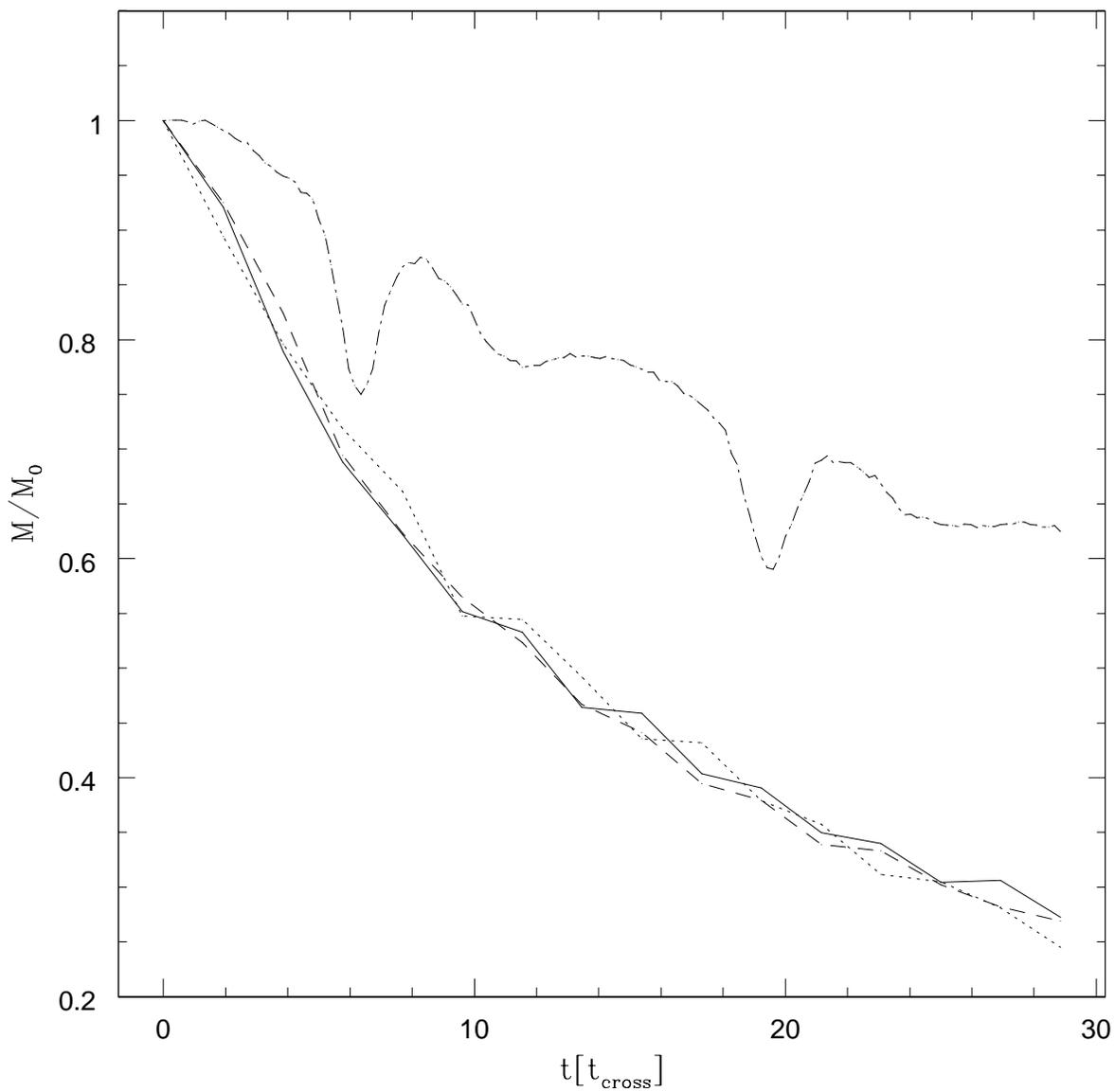}
\caption{Stellar mass belonging to the cluster, in units of the initial cluster mass, as a
function of time, expressed in units of the bulge crossing time. Solid line:
quasi-circular orbit with $N=1.6\times10^5$ particles. Dashed line: quasi-circular orbit
with $N=1.6\times10^4$ particles. Dotted line: loop orbit with $e=0.27$.
Dot-dashed: loop orbit with $e=0.57$. \label{mloss}}
\end{figure}

The limiting radius $r_{L}$ is then defined as the radius of the sphere (centered in the cluster
density center) in which the cluster `emerges' from the stellar background.
In Fig.\ref{mloss}, the
evolution of the cluster mass, expressed in units of the initial mass $M_{0}$, is shown versus time
for all the four simulations performed. \\
In the case of a cluster moving on orbits with apocenter $\leq 3.5 r_b$, the mass loss
is dramatic: after about 30 $t_{\rm{cross}}$ the cluster loses  about 75$\%$ of its mass;
the best fit of the mass evolution as a function of time is given by:\\
\begin{equation}
\frac{M(t)}{M(0)}=0.77e^{-t/12}+0.21,
\end{equation}
where $t$ is expressed in units of the bulge crossing time $t_{cross}$.
In the remaining case, when the orbit extends up to 7.5 $r_b$, the mass loss rate considerably
diminishes and the cluster mass, after 30 $t_{\rm{cross}}$, is still about 60$\%$ of its initial value.
As is evident in Fig.\ref{mloss}, in this case the mass loss rate increases every time the cluster
passes at the minimum distance from the galaxy center and not all particles which become unbound at
perigalacticon are still so  while moving again to apogalacticon. It is possible to point out a
region around the galaxy center inside which the cluster suffers more of mass loss:
in our case (cluster concentration equal to 1.1) this region corresponds roughly
to $r\leq$4 $r_b$. This conclusion is accordance to what found in \citet{mioc}, where great
mass loss occurs for clusters with comparable central density moving on quasi--radial
orbits within such region.\\
Finally, we want to stress that the choice of performing some of the simulations with a
reduced number of particles ($N=1.6\times10^4$) did not affect the mass loss rate over
the time interval of 30 $t_{\rm{cross}}$. Actually, Fig.\ref{mloss} clearly shows that,
for the cluster in a quasi-circular orbit, the mass loss rate is the same using either
$N=1.6\times10^5$ (solid line) or a ten times smaller $N$ (dashed line).

\subsection{Mass segregation and mass function}\label{massf}
\begin{figure}
\plotone{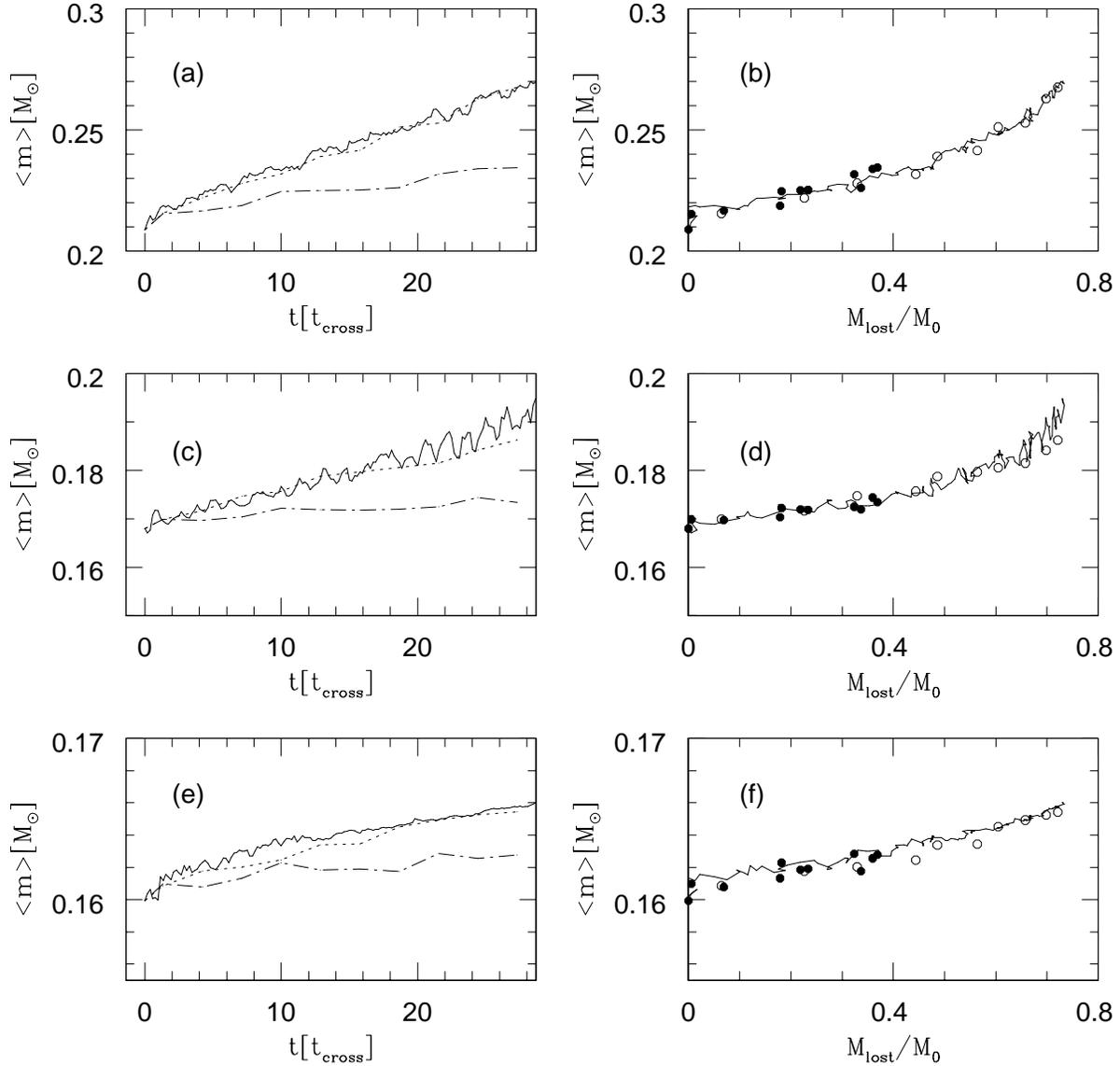}
\caption{Left column: time evolution of the mean mass of stars in three different
regions of space centered with the cluster. Panel (a): Mean stellar mass inside
$r=0.016 r_b$, in the case of quasi-circular orbit (solid line), loop orbit with $e=0.27$
(dashed line) and loop orbit with $e=0.57$ (dot-dashed). Panel (c): Mean stellar mass
between $r=0.016 r_b$ and $r=0.036 r_b$. Panel (e): Mean stellar mass outside $r=0.036 r_b$.
Right column: evolution of the mean mass of stars in three different regions of space as a
function of the fraction of mass lost from the system (in this case both the mass lost
and the mean stellar mass have been averaged on time interval of 2.9 $t_{cross}$ for
the two loop orbits with greater ellipticities). Panel (b): Mean stellar mass inside
$r=0.016 r_b$, in the case of quasi-circular orbit (solid line), loop orbit with
$e=0.27$ (open circles) and loop orbit with $e=0.57$ (solid circles). Panel (d):
Mean stellar mass between $r=0.016r_b$ and $r=0.036r_b$. Panel (f): Mean stellar
mass outside $r=0.036r_b$.  \label{msegreg-mgc}}
\end{figure}

\begin{figure}
\plotone{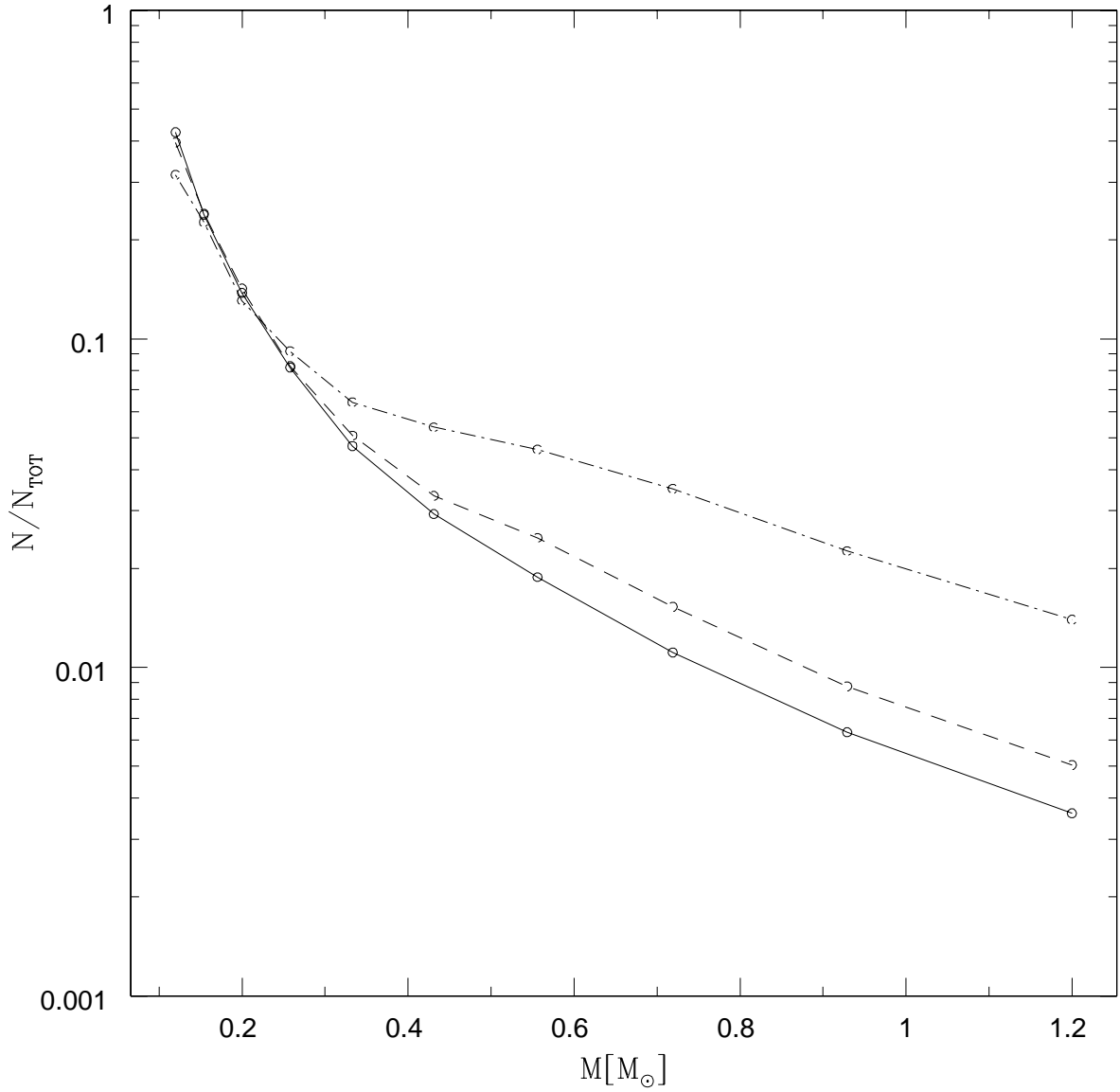}
\caption{Mass function of stars belonging to the cluster at three different epochs:
when the cluster has lost the $20\%$ (solid line), the $35\%$ (dashed line) and
the $75\%$ of its initial cluster mass (dot-dashed line). Only the evolution of
the mass function in the case of quasi-circular orbit is shown, because the
other curves (corresponding to orbits with greater ellipticities) coincide
with these.\label{fmassatot}}
\end{figure}

As explained in Sect.\ref{Cluster1}, we adopted mass segregation in the globular cluster initial conditions,
for we aim at simulating a dinamically evolved cluster, whose orbit had decayed in the inner galactic
region due to dynamical friction. As the cluster begins to lose
stars, the distribution of stars of different masses in the system starts evolving.
This is shown
in the left column of Fig.\ref{msegreg-mgc}, where the mean mass of stars populating three different spatial
regions versus time is plotted. The first region corresponds to the sphere with radius $r=0.016r_b$
(corresponding to $r=3.22$ pc, with our choice of $M_b$ and $r_b$) centered on the cluster, which initially
contains $40\%$ of the total mass of the system; the second region is the spherical shell with inner and
outer radius $r=0.016r_b$ and $r=0.036r_b$ ($r=7.18$ pc), which initially contains $80\%$ of the cluster mass;
the third region is that outside $r=0.036r_b$.
As time passes by, low mass stars begin to escape from the system,
and the mean stellar mass in the two inner regions starts to rise, while the external one remains quite constant.
The increasing of the mean stellar mass versus time in the central cluster regions is particularly evident in the case of the quasi-circular orbit and of the loop with ellipticity $e=0.27$, because of the greater mass loss in these two cases.  Plotting the mean stellar mass in the three regions defined above as a function of the fraction of mass lost, we see (Fig.\ref{msegreg-mgc}, right column) that the evolution of the mean mass depends mostly on the fraction of mass lost from the system rather than on the number of stars populating the cluster and on the cluster orbital path.\\
The differential mass loss obviously influences the shape of the mass function at different times. In Fig.\ref{fmassatot} the mass function of stars belonging to the cluster is shown
at three different epochs, when the cluster has lost respectively the $20\%$, the $35\%$ and the $75\%$ of
its initial mass.  As the cluster loses stars in the galactic field, the mass function evolves towards
flatter configurations, because of the preferential loss of low-mass stars, that, accordingly to the initial
mass segregation, are located mostly in the external regions of the cluster.
The evolution of the mass function appears to be driven by the fraction of mass loss, rather than by other
parameters (like the total number of stars in the system and the orbital type of the parent cluster) confirming
the findings of \citet{bm03}. This is evident from the fact that, for a given fraction of mass lost, the
curves found for the different orbits in practice coincide.

\section{Conclusions}
The main results of our work may be resumed as follows:
\begin{enumerate}
\item Stars are lost from the system along a direction which results from the composition of the
direction towards the galactic center and the cluster velocity around the galaxy, thus leading to
the peculiar S-shape found in the outermost region of the cluster. Once formed,
tidal tails are elongated such to remain parallel to the cluster orbit, with a trailing tail that lies
slightly inside the orbit and a leading tail slightly outside it. Tails are excellent tracers of the
cluster orbit near the pericenter, while, at the apocenter, they tend to deviate from the orbital path.
\item Tidal tails have a clumpy structure which cannot be associated with an episodic mass loss or tidal
shocks with galactic compact sub-structures, since stars are lost from the cluster continuously and since
the interaction with the bulge  is not episodic. These clumps are not bound self-gravitating systems,
they are rather due to a local deceleration of the motion of the stars along the tails.
\item The observational evidence found for M15 and $\omega$ Centauri that the velocity dispersion
increases and then remains constant at large radii is explained in terms of the so--called `tidal heating':
the stars that evaporate outside the tidal radius of the cluster undergo mainly the interaction with
the external field, thus acquiring the higher velocity dispersion pertaining to that field.
\end{enumerate}

\section{Acknowledgement}
Part of this work has been done using the IBM SP4 platform located at CINECA, thanks to the grant
\emph{inarm007} obtained in the framework of INAF-CINECA agreement (http://inaf.cineca.it).\\
The authors are greatly thankful to dr. A. Vicari (Univ. of Rome La Sapienza, Italy) and dr. G. Carraro
(Univ. of Padova, Italy) for their helpful suggestions.

\appendix

\section{Adimensionalization of the equations}\label{scal}
The equations of motion of the $j-$th star of the cluster, interacting with all the other cluster
members and with
the bulge are:
\begin{equation}
\ddot{\bf r}_j = \sum_{i=1}^N\frac{Gm_i}{r_{ij}^3}\left({\bf r}_i-{\bf r}_j\right)+
\nabla U_b\mid_{(x_j,y_j,z_j)} \label{xdot}
\end{equation}

where ${\bf r}_j=(x_j,y_j,z_j)$ is the position vector, $r_{ij}$ is the distance between the $i-$th and the
$j-$th particle and $U_b$ is the bulge potential. In the case of
the Schwarzschild model \citep{scw}:
\begin{eqnarray}
U_b(x,y,z)&=&4\pi GM_b\left[-\frac{1}{r}\ln\left(\frac{r}{r_b}+\sqrt{1+(r/r_b)^2}\right)+c_1\frac{3z^2-r^2}{2r_b^3
\left(1+c_2(r/r_b)^2\right)^{3/2}}+\right.\nonumber \\
& & \left.-3c_3\frac{x^2-y^2}{r_b^3\left(1+c_4(r/r_b)^2\right)^{3/2}} \right] \label{rpot},
\end{eqnarray}
where $r_b$ and $M_b$ are the bulge core radius and the bulge mass respectively.
Equation (\ref{rpot}) can be rewritten as a product of a dimensional factor and a dimensionless function as:
\begin{eqnarray}
 U_b(x',y',z')&=&4\pi \frac{GM_b}{r_b}\left[-\frac{1}{r'}\ln\left(r'+\sqrt{1+r'^2}\right)+c_1\frac{3z'^2-r'^2}
{2(1+c_2r'^2)^{3/2}}+\right.\nonumber \\
& & {} \left.-3c_3\frac{x'^2-y'^2}{(1+c_4r'^2)^{3/2}} \right] \label{rprimepot}
\end{eqnarray}
where
\begin{equation}
r'=\frac{r}{r_b}, \ \ x'=\frac{x}{r_b}, \ \ y'=\frac{y}{r_b}, \ \ z'=\frac{z}{r_b},
\end{equation}
Also the first term on the right side of Eq.(\ref{xdot}) can be written as the product of a dimensional factor and a
dimensionless one:
\begin{equation}
\sum_{i=1,}^N\frac{Gm_i}{r_{ij}^3}\left({\bf r}_i-{\bf r}_j\right)=G\frac{M_b}{r_b^2}\sum_{i=1,}^N\frac{m'_i}{{r'_{ij}}^3}\left({\bf r}'_i-{\bf r}'_j\right),
\end{equation}
with $m'_i=m_i/M_b$.\\
Finally, once defined a dimensionless time
\begin{equation}
t'=\frac{t}{t_{cross}},
\end{equation}
being $t_{cross}=(r_b^3/GM_b)^{1/2}$ the bulge crossing time, Eq.\ref{xdot} may be written as:
\begin{equation}
\frac{d^2{\bf r}_i'}{dt'^2}=\sum_{i\ne j}^N \frac{Gm'_i}{\left(r'_{ij}\right)^3}\left({\bf r}_i'-{\bf r}_j'\right)+4\pi \nabla U'. \label{new}
\end{equation}
This implies that, once assigned the initial conditions ${\bf r}_i'(0)$, ${\bf v}_i'(0)$,
the existence of a unique solution for the Eq. (\ref{new}) ensures that all the results obtained can be scaled in terms of
the ratios $r/r_b$, $m/M_b$ and $t/t_{cross}$.

\section{The construction of multi--mass King model}\label{kingmod}
As described in \citet{king66}, in a single-mass isotropic King model the phase--space stellar distribution function is given by:
\begin{eqnarray}
f(r,v)&=&\alpha\left[\exp\left(-\frac{E}{m\sigma^2}\right)-\exp\left(\frac{C}{\sigma^2}\right)\right],
\mbox{ if } E\leq -mC  \label{distrfun}\\
{}&=& 0 \mbox{ otherwise} \nonumber
\end{eqnarray}
where
\begin{equation}
E=\frac{1}{2}mv^2+m\psi(r)
\end{equation}
is the energy of a star, $\psi(r)$ is the mean gravitational potential generated by
the cluster and $\alpha$ is a normalization constant. The `global' parameter $C$ is related to the
tidal radius $r_t$ by the implicit relation
\begin{equation}\label{tid}
\psi(r_t)+C=0.
\end{equation}
The mass density can be found integrating the distribution function $f(r,v)$ over the velocity,
obtaining an explicit relation for $\rho$ as a function of the potential $\psi$:
\begin{eqnarray}
\rho&=&\int_{E\leq-mC}f(r,v)4\pi v^2dv=\\
{}&=&4\pi m\alpha e^{C/\sigma^2}(2\sigma^2)^{3/2}\left[-\frac{1}{2}\left(-\frac{\psi+C}{\sigma^2}\right)^{1/2}+
\right. \nonumber \\
{} &+&\left.\frac{\sqrt\pi}{4}\exp\left(-\frac{\psi+C}{\sigma^2}\right)\mbox{erf}\left(\sqrt{-\frac{\psi+C}{\sigma^2}}\right)-
\frac{1}{3}\left(-\frac{\psi+C}{\sigma^2}\right)^{3/2}\right]=\\
{}&=&4\pi m\alpha e^{C/\sigma^2}(2\sigma^2)^{3/2}\left[-\frac{1}{2}\sqrt{-U}+\right. \nonumber \\
{} &+&\left.\frac{\sqrt\pi}{4}\exp\left(-U\right)\mbox{erf}\left(\sqrt{-U}\right)-
\frac{1}{3}\left(-U\right)^{3/2}\right]=\\
{}&\equiv&k\tilde\rho\left(U\right)
\end{eqnarray}
where $U\equiv(\psi+C)/\sigma^2$ is the dimensionless potential,
$k=4\pi m\alpha e^{C/\sigma^2}(2\sigma^2)^{3/2}$
and $\tilde\rho$ is the dimensionless density, which explicitely depends only on $U$.
Once assigned initial conditions for the potential $\psi$ and its derivative $\psi'$ in $r=0$, the Poisson equation
\begin{displaymath}
\left\{\begin{array}{ll}
d^2 \psi/dr^2=4 \pi G \rho,\\
\psi(0)=\psi_0 \nonumber \\
\psi'(0)=0 \nonumber
\end{array}\right.
\end{displaymath}
can be rewritten in terms of the dimensionless potential, in the form:
\begin{displaymath}\label{poieq}
\left\{\begin{array}{ll}
d^2 U/d\tilde r^2=9\rho(U)/\rho(0)=9\tilde\rho(U)/\tilde\rho(0) \\
U(0)=U_0 \nonumber \\
U'(0)=0
\end{array}\right. \label{poisson}
\end{displaymath}
where $\rho(0)=\rho(U_0)$ and $\tilde r= r/r_{core}$ being
\begin{equation}
r_{core}^2=\frac{9\sigma^2}{4\pi G \rho_0} \label{B8}
\end{equation}
the King radius.

Once assigned as initial parameters $U_0$, $U'_0$, the Poisson equation can be integrated,
obtaining the dimensionless potential $U(\tilde r)$, the dimensionless mass density $\tilde\rho(\tilde r)$
and the tidal radius $\tilde r_t$, being $\tilde r_t=r_t/r_{core}$ with $r_{core}$ yet not determined.
To determine the core radius $r_{core}$ and the costant $k$ (which depends, among others, on the normalization constant $\alpha$), it is possible to procede as follows.\\
Once assigned as initial parameters the total mass of the cluster $M_{tot}$ and the velocity dispersion $\sigma$ of stars in the system, using the following relations\\
\begin{eqnarray}\label{mtot}
M_{tot}&=&\int_0^{r_t}4\pi\rho(r)r^2dr= \nonumber\\
& &=4\pi kr_{core}^3\int_0^{\tilde r_t}\tilde \rho(\tilde r)\tilde r^2d\tilde r
\end{eqnarray}
and the Eq.\ref{B8},
it is possible to  calculate $r_{core}$ and $k$ and hence to obtain $\rho(r)$, $r_{core}$ and $r_t=
\tilde r_t r_{core}$.

In a multi--mass isotropic King model, as described in \citet{costa76}, stars are first grouped in $n$ different mass
classes, each characterized by a mass $m_{i}$. The phase--space stellar distribution function for the $i$-th mass class is given by:
\begin{eqnarray}
f_{i}(r,v)&=&\alpha_i\left[\exp\left(-\frac{E_i}{m_i\sigma_i^2}\right)-\exp\left(\frac{C}{\sigma_i^2}\right)\right],
\mbox{ if } E_i\leq -m_iC  \label{distrfuni}\\
{}&=& 0 \mbox{ otherwise} \nonumber
\end{eqnarray}
where
\begin{equation}
E_i=\frac{1}{2}m_iv^2+m_i\psi(r)
\end{equation}
is the energy of a star in the $i$-th mass class, $\psi(r)$ is the mean gravitational potential generated by
the \emph{\/whole} cluster, $\alpha_i$ is a normalization constant and $C$ is related to the cluster tidal radius $r_t$ by Eq.\ref{tid}.  
Once again, the mass density for the $i$-th mass class can be found integrating the distribution function $f_i(r,v)$ over velocities,
obtaining an explicit relation for $\rho_i$ in terms of the dimensionless potential $U$ defined above and
the ratio $\sigma^2/\sigma_i^2$:
\begin{eqnarray}
\rho_i&=&\int_{E_i\leq-m_iC}f_i(r,v)4\pi v^2dv=\\
{}&=&4\pi m_i\alpha_i e^{C/\sigma_i^2}(2\sigma_i^2)^{3/2}\left[-\frac{1}{2}\left(-\frac{\psi+C}{\sigma_i^2}\right)^{1/2}+
\right. \nonumber \\
{} &+&\left.\frac{\sqrt\pi}{4}\exp\left(-\frac{\psi+C}{\sigma_i^2}\right)\mbox{erf}\left(\sqrt{-\frac{\psi+C}{\sigma_i^2}}\right)-
\frac{1}{3}\left(-\frac{\psi+C}{\sigma_i^2}\right)^{3/2}\right]=\\
{}&=&4\pi m_i\alpha_i e^{C/\sigma_i^2}(2\sigma_i^2)^{3/2}\left[-\frac{1}{2}\sqrt{-U\frac{\sigma^2}{\sigma_i^2}}+\right. \nonumber \\
{} &+&\left.\frac{\sqrt\pi}{4}\exp\left(-U\frac{\sigma^2}{\sigma_i^2}\right)\mbox{erf}\left(\sqrt{-U\frac{\sigma^2}{\sigma_i^2}}\right)-
\frac{1}{3}\left(-U\frac{\sigma^2}{\sigma_i^2}\right)^{3/2}\right]=\\
{}&\equiv&k_i\tilde\rho_i\left(U,\sigma^2/\sigma_i^2\right)
\end{eqnarray}
where 
$k_i=4\pi m_i\alpha_i e^{C/\sigma_i^2}(2\sigma_i^2)^{3/2}$
and $\tilde\rho_i$ is the dimensionless density.
Note that the density profiles $\rho_i$ of the $i-th$ mass class are related to the ``global'' density distribution $\rho$ by the relation:
\begin{equation}\label{sum}
 \rho(r)=\sum_{i=1}^n\rho_i(r).
\end{equation}

To distribute stars in the cluster according to this isotropic multimass King model, we proceeded in the following way:
\begin{itemize}
\item Once assigned $U_0$, $U'_0$, the total cluster mass $M_{tot}$ and the velocity dispersion $\sigma$, we integrate the
Poisson equation as in the case of a sigle mass model previously described, obtaining the
dimensionless potential $U(\tilde r)$,
the ``global'' mass density $\rho$, the core radius $r_{core}$ and the tidal radius $r_t$.
\item Then we assigned the mass $m_i$ of stars in the $i$-th mass class and the total mass
$M_{tot,i}$ of each mass class (i.e. $M_{tot,i}=n_i\times m_i$, being $n_i$ the number
of stars populating the $i$-th mass class). In our case, we chose to set the masses
$M_{tot,i}$ according to the Salpeter's mass function.
For a given value of the ratio $\sigma^2/\sigma_1^2$ (once obtained all the other values according to energy equipartition using the relation $m_1\sigma_1^2=m_i\sigma_i^2$ for $i\geq 2$), we calculate $\tilde \rho_i(U,\sigma^2/\sigma_i^2)$
and then the coefficients $k_i$ using Eq.\ref{mtot} applied to the $i-th$ mass class.
\item We varied the ratio $\sigma^2/\sigma_1^2$ until the relation \ref{sum} is satisfied with the desired accuracy.
\end{itemize}
Finally, stars velocities were generated according to Eq.\ref{distrfuni}.

\section{The GC initial mass function}\label{remn}
The GC we considered in our simulations is supposed to have an age of $t_{gc}\sim 10^{9.5}$ yr,
thus only stars more
massive than $\sim 1.2$ M$_\odot$ are at that time evolved up to a compact remnant
\citep{scl97,dom99}.
For this reason, we considered only stars distributed according to the Salpeter's
MF with masses in the range $0.12 \leq m \leq 1.2$ M$_\odot$.
Moreover, we assumed that the contribution to the low mass population due to the
mentioned remnants is practically negligible.

Indeed, according to the Salpeter's MF, the ratio between the number of
remnants whose progenitor had a mass around $m_p$ and the number of stars
with mass around $m$ is
\begin{equation}
\frac{N_{\rm remn}}{N_m}=\left(\frac{m}{m_p}\right)^{2.35}. \label{nremn}
\end{equation}
Supposing that such progenitors are those giving rise to remnants with
mass $m$, then, from the estimates in \cite{scl97} and \cite{dom99},
$m_p ($M$_\odot) \simeq 9.5 (m - 0.45)$, with $m>0.45$ M$_\odot$ because stars
with lower mass remnants cannot be evolved in a Hubble time.
Thus, substituting in Eq. (\ref{nremn}),
\begin{eqnarray}
\frac{N_{\rm remn}}{N_m}&\simeq &\left(\frac{0.11 \times m}{m-0.45}\right)^{2.35}
\mbox{ if } m\geq m_l, \label{ratio} \\
{}&\simeq & 0 \mbox{ otherwise} \nonumber
\end{eqnarray}
where $m_l$ is the lowest mass a remnant can have at the assumed cluster age. From
fitting the above--cited estimates, this lower limit turns out to be $m_l\sim
(0.45\log t_{gc} - 1.2)^{-3.1}+0.45 \sim 0.59$ M$_\odot$.

One can see that the ratio in Eq. (\ref{ratio}) is monotonically decreasing for $m\geq m_l$,
hence the maximum takes place for the lowest mass class we used in the model, i.e. $m=0.71$
M$_\odot$, for which $N_{\rm remn}/N_{0.71} \sim 0.05$. Since in our
numerical representation $N_{0.71}/N \simeq 0.01$ ($N$ is the total number of particles),
then in this class there should have been about $5\times 10^{-4}N$ remnants.
Thus, bearing in mind that the less populated mass class contains $\sim 2\times 10^{-3}N$,
we can reasonably affirm that the MF we assumed for the initial conditions was not
substantially affected by stellar evolution neither at the initial time $t_{gc}$ nor
later during the simulation (because it lasts much shorter than $t_{gc}$).


\clearpage

\clearpage
\clearpage

\clearpage
\clearpage

\clearpage
\clearpage

\clearpage

\clearpage

\clearpage

\clearpage

\clearpage

\clearpage

\clearpage

\clearpage

\clearpage

\clearpage

\clearpage

\clearpage
\begin{deluxetable}{cccc}
\tablecaption{Orbital parameters. \label{tbl-1}}
\tablewidth{0pt}
\tablehead{
\colhead{N}   & \colhead{$e$} &
\colhead{$(x_0,y_0,z_0)$} & \colhead{$(v_{x,0},v_{y,0},v_{z,0})$}
}
\startdata
$1.6\times10^5$ &0.03  &(0,0,2.50) &(0,1.82,0)\\
$1.6\times10^4$ &0.03  &(0,0,2.50) &(0,1.82,0)\\
$1.6\times10^4$ &0.27  &(0,0,3.50) &(0,1.30,0)\\
$1.6\times10^4$ &0.57  &(0,0,7.50) &(0,0.78,0)\\
\enddata
\tablecomments{Orbital parameters of the cluster in the four simulations performed.
The first column shows the total number of particles used in each simulation.
Initial positions and velocities of the cluster with respect to the galaxy
center (columns 3 and 4) have been expressed, respectively, in units of the
galaxy bulge radius $r_b$ and of the bulge typical velocity dispersion
$r_b/t_{\rm{cross}}$}

\end{deluxetable}

\clearpage
\begin{deluxetable}{ccccccc}
\tablecaption{Clumps emersion from the background. \label{tbl-2}}
\tablewidth{0pt}
\tablehead{
\colhead{$r[r_b]$}& & & \colhead{$\rho_{cl}/\rho_{\ast}$}& & &\colhead{$t[t_{cross}]$}
}
\startdata
2.04& & & $\leq 0.1$& & & 9.62 \\
3.44& & & 0.3       & & & 13.5 \\
3.47& & & 0.3       & & & 19.2 \\
2.12& & & $\leq 0.1$& & & 23.1 \\
3.44& & & 0.3       & & & 25.0 \\
2.05& & & $\leq 0.1$& & & 28.8 \\
\enddata
\tablecomments{Clumps emersion from the background density, in the case of loop orbit with ellipticity $e=0.27$. In the first column the cluster distance from the galaxy center is given; the second column shows the ratio between the clumps local density and that of the stellar background; the third column shows the time from the beginning of the simulation, expressed in units of the bulge crossing time.}

\end{deluxetable}


\begin{thebibliography}{}
\bibitem[Aarseth(1985)]{aars}Aarseth, S.J. 1985, in `Multiple time scales',
 Acad. Press, 378
\bibitem[Barnes \& Hut(1986)]{bh}Barnes, J. \& Hut, P. 1986, \nat, 324, 446
\bibitem[Baumgardt \& Makino(2003)]{bm03}Baumgardt, H., Makino, J. 2003, \mnras, 340, 227
\bibitem[Bertola et al.(1991)]{ber91}Bertola, F., Vietri, M., Zeilinger, W. W. 1991, \apj, 374, L13
\bibitem[Binney and Tremaine(1987)]{BT}Binney, J., Tremaine, S. 1987,
{\it Galactic Dynamics}, Princeton Univ. Press (Princeton, USA)
\bibitem[Capuzzo Dolcetta \& Donnarumma(2001)]{imma01}Capuzzo Dolcetta, R., Donnarumma, I. 2001, \mnras, 328, 645
\bibitem[Capuzzo Dolcetta \& Tesseri(1997)]{cdt97}Capuzzo Dolcetta, R., Tesseri, A. 1997, \mnras, 292, 808
\bibitem[Capuzzo Dolcetta \& Tesseri(1999)]{cdt99}Capuzzo Dolcetta, R., Tesseri, A. 1999, \mnras, 308, 961
\bibitem[Capuzzo Dolcetta \& Vicari(2003)]{cv03}Capuzzo Dolcetta, R., Vicari, A. 2004, submitted to \mnras
(astro-ph/0309488)
\bibitem[Capuzzo Dolcetta \& Vignola(1997)]{cv97}Capuzzo Dolcetta, R., Vignola, L. 1997, \aap, 327, 130
\bibitem[Carraro \& Lia(2000)]{car00}Carraro, G., Lia, C. 2000, \aap, 357, 977
\bibitem[Casertano \& Hut(1985)]{ch85}Casertano, S., Hut, P. 1985,  \apj, 298, 80
\bibitem[Cohn(1980)]{cohn80}Cohn, H. 1980, \apj, 242, 765
\bibitem[Combes et al.(1999)]{combes99}Combes, F., Leon, S, Meylan, G. 1999, \aap, 352, 149
\bibitem[Da Costa \& Freeman(1976)]{costa76}Da Costa, G. S., Freeman, K. C. 1976, \apj, 206, 128
\bibitem[de Zeeuw \& Merritt(1983)]{zeeuw}de Zeeuw, T., Merritt, D. 1983, \apj, 267, 571
\bibitem[Dehnen et al.(2004)]{dehn04}Dehnen, W., Odenkirchen, M., Grebel, E. K., Rix, H. W. 2004, \aj,
127, 2753
\bibitem[Di Matteo et al.(2004)]{dm04}Di Matteo, P., Capuzzo Dolcetta, R., Miocchi, P., 2004, submitted to Celestial Mechanics and Dynamical Astronomy
\bibitem[Djorgovski et al.(1993)]{dpc93}Djorgovski, S., Piotto, G., Capaccioli, M. 1993, \aj, 105, 2148\bibitem[Dominguez et al (1999)]{dom99}Dominguez, I., Chieffi, A., Limongi, M. \& Straniero, O. 1997,
\apj, 524-1, 226
\bibitem[Drukier et al.(1998)]{druk}Drukier, G. A., Slavin, S. D., Cohn, H. N., Lugger, P. M., Berrington, R. C., Murphy, B. W., Seitzer, P. O. 1998, \aj, 115, 708
\bibitem[Fall \& Zhang(2001)]{fz01}Fall, S. M., Zhang, Q. 2001, \apj, 561, 751
\bibitem[Grillmair et al.(1986)]{grill86}Grillmair C., Pritchet C., van de Bergh S. 1986, \aj, 91, 1328
\bibitem[Grillmair et al.(1995)]{grill95}Grillmair, C. J., Freeman, K. C., Irwin, M., Quinn, P. J. 1995, \aj, 109, 2553
\bibitem[H\`enon(1961)]{hen61}H\'enon, M.  1961, Ann.d'Ap., 24, 369
\bibitem[Hernquist \& Katz(1989)]{bib5} Hernquist, L. \& Katz, N. 1989, \apjs, 70, 419
\bibitem[Heggie \& Hut(2003)]{hh03} Heggie, D. C., Hut, P. 2003, {\it The Gravitational Million-Body Problem}, Cambridge Univ. Press (Cambridge, UK)
\bibitem[King(1966)]{king66}King, I. R.  1966, \aj, 71, 276
\bibitem[Lehmann \& Scholz(1997)]{ls97}Lehmann, I., Scholz, R.D. 1997, \aap, 320, 776
\bibitem[Lee et al.(2003)]{lee03}Lee, K. H., Lee, H. M., Fahlman, G. G., Lee, M. G. 2003, \aj, 126, 815
\bibitem[Leon et al.(2000)]{lmc00}Leon, S., Meylan, G., Combes, F. 2000, \aap, 359, 907
\bibitem[McLaughlin(1995)]{mcl95}McLaughlin D. E. 1995, \aj, 109, 2034
\bibitem[Miocchi \& Capuzzo-Dolcetta(2002)]{bib2}Miocchi, P. \&
Capuzzo-Dolcetta, R. 2002, \aap, 382, 758
\bibitem[Miocchi et al.(2004)]{mioc}Miocchi, P., Capuzzo-Dolcetta, R., Di Matteo, P., \& Vicari, A.
2004, in preparation.
\bibitem[Murali \& Weinberg(1997a)]{mw97a}Murali, C., Weinberg, M. D. 1997a, \mnras, 291, 717
\bibitem[Murali \& Weinberg(1997b)]{mw97b}Murali, C., Weinberg, M. D. 1997b, \mnras, 288, 749
\bibitem[Odenkirchen et al.(2001)]{oden01}Odenkirchen, M., Grebel, E. K., Rockosi, C. M., Dehnen, W., Ibata, R., Rix, H. W., Stolte, A., Wolf, C., Anderson, J. E. Jr., Bahcall, N. A., Brinkmann, J., Csabai, I., Hennessy, G., Hindsley, R. B., Ivezic, Z., Lupton, R. H., Munn, J. A., Pier, J. R., Stoughton, C., York, D. G. 2001, \apj, 548, L165
\bibitem[Odenkirchen et al.(2002)]{oden02}Odenkirchen, M., Grebel, E. K., Dehnen, W., Rix, H. W., Cudworth, K. M. 2002, \aj, 124, 1497
\bibitem[Odenkirchen et al.(2003)]{oden03}Odenkirchen, M., Grebel, E, K., Dehnen, W, Rix, H. W., Yanny, B., Newberg, H. J., Rockosi, C. M., Mart\'inez-Delgado, D., Brinkmann, J., Pier, J. R. 2003, \aj, 126, 2385
\bibitem[Ostriker(1985)]{ostr85}Ostriker, J. P. 1985, Dynamics of
    Star Clusters, J. Goodman and P. Hut, Dordrecht: Reidel, 347
\bibitem[Pesce et al.(1992)]{pesce}Pesce, E., Capuzzo-Dolcetta, R., Vietri, M. 1992, \mnras, 254, 466
\bibitem[Salpeter (1955)]{salp}Salpeter, E. E. 1955, \apj, 121, 161
\bibitem[Scarpa et al.(2003)]{scarpa}Scarpa, R., Marconi, G., Gilmozzi, R. 2003, \aap, 405, L15
\bibitem[Schwarzschild(1979)]{scw}Schwarzschild, M. 1979, \apj, 232, 236
\bibitem[Shapley(1918)]{shapley}Shapley, H. 1918, PASP, 30, 42
\bibitem[Siegel et al.(2001)]{sieg00}Siegel, M. H., Majewski, S. R., Cudworth, K. M., Takamiya, M. 2001, \aj, 121, 935
\bibitem[Straniero et al (1997)]{scl97}Straniero, O., Chieffi, A. \& Limongi, M. 1997, \apj, 490, 425.
\bibitem[Testa et al.(2000)]{testa00}Testa, V., Zaggia, S. R., Andreon, S., Longo, G., Scaramella, R., Djorgovski, S. G., de Carvalho, R. 2000, \aap, 356, 127

\end{thebibliography}
\end{document}